\documentclass[useAMS,usenatbib,onecolumn]{mnras}
\usepackage{amsmath,bm,amssymb}
\usepackage{dcolumn}
\usepackage[utf8]{inputenc}
\usepackage{graphics,graphicx}
\usepackage{float}
\usepackage{threeparttable}
\usepackage{url}
\usepackage{epstopdf}
\usepackage{siunitx}
\usepackage{epstopdf}
\usepackage{booktabs}

\usepackage{longtable}

\usepackage{color}

\newcommand{\green}[1]{{\color{green} #1}}

\newcommand{\cm}{cm$^{-1}$}

\newcommand{\bra}[1]{\langle #1|}
\newcommand{\ket}[1]{|#1\rangle}

\newcommand{\Duo}{{\sc Duo}}

\newcommand{\ai}{\textit{ab initio}}

\newcommand{\X}{$X\,{}^{1}\Sigma^{+}$}
\newcommand{\A}{$A\,{}^{1}\Pi$}
\newcommand{\C}{$C\,{}^{1}\Sigma^{-}$}
\newcommand{\D}{$D\,{}^{1}\Delta$}
\newcommand{\E}{$E\,{}^{1}\Sigma^{+}$}
\newcommand{\as}{$a\,{}^{3}\Sigma^{+}$}
\newcommand{\bp}{$b\,{}^{3}\Pi$}
\newcommand{\dd}{$d\,{}^{3}\Delta$}
\newcommand{\es}{$e\,{}^{3}\Sigma^{-}$}

\newcommand{\XA}{\A\--\X}
\newcommand{\AX}{\A\--\X}
\newcommand{\EX}{\E\--\X}
\newcommand{\XE}{\E\--\X}
\newcommand{\XX}{\X\--\X}

\newcommand{\pn}{$^{31}$P$^{14}$N}

\usepackage{graphicx}

\graphicspath{ {img/} }

\newcommand{\name}{PaiN}

\title[ExoMol line lists -- {LXIV}. PN]{ExoMol line lists -- LXIV: Empirical rovibronic spectra of phosphorous mononitride (PN) covering the IR and UV regions }

\author[M. Semenov et al.]{
Mikhail Semenov$^{1}$, Nayla El-Kork$^{2}$, Sergei  N. Yurchenko$^{1}$, Jonathan Tennyson$^{1}$\thanks{The corresponding author:
j.tennyson@ucl.ac.uk}
\vspace*{4mm}\\
$^{1}$Department of Physics and Astronomy, University College London, Gower Street, WC1E 6BT London, United Kingdom\\
$^{2}$Space and Planetary Science Center, Khalifa University, Abu-Dhabi, UAE.\\
}

\date{Accepted XXX. Received YYY; in original form ZZZ}

\pubyear{2024}

\begin{document}

\label{firstpage}

\pagerange{\pageref{firstpage}--\pageref{lastpage}}

\maketitle

\begin{abstract}

A new phosphorous mononitride (\pn, ${}^{31}$P${}^{15}$N) line list PaiN covering infrared, visible and ultraviolet regions is presented. The PaiN line list extending to the$A\,{}^{1}\Pi$ --  $X\,{}^{1}\Sigma^{+}$ vibronic band system, replaces the previous YYLT ExoMol line list for PN. A thorough analysis of high resolution experimental spectra from the literature involving the $X\,{}^{1}\Sigma^{+}$ and $A\,{}^{1}\Pi$ states is conducted, and many perturbations to the $A\,{}^{1}\Pi$ energies are considered as part of a comprehensive MARVEL study. {\it Ab initio} potential energy and coupling curves from the previous work [Semenov et al., Phys. Chem. Chem. Phys., 23, 22057 (2021)] are refined by fitting their analytical representations to 1224 empirical energy levels determined using the MARVEL procedure. The PaiN line list is compared to previously observed spectra, recorded and calculated lifetimes, and previously calculated partition functions. The {\it ab initio} transition dipole moment curve for the  $A$--$X$ band is scaled to match experimentally measured lifetimes. The line list is suitable for temperatures up to 5~000~K  and wavelengths longer than 121 nm. PaiN is available from \url{www.exomol.com}.

\end{abstract}




\section{Introduction}

Phosphorus mononitride (PN) has emerged as a molecule of significant astrophysical interest, challenging earlier theoretical predictions about the abundance of phosphorus-bearing
species in space. While phosphorus monoxide (PO) was initially expected to be the most prevalent diatomic phosphorus molecule based on early lab experiments \citep{83ThAnPr.PN, 87MiBeHe.PN},
observations have consistently found PN to be more widespread across various astronomical environments \citep{Fontani24, Rivilla16}, albeit with a lower abundance. PN has been detected in star-forming regions like the Orion Nebula \citep{87TuBaxx.PN, 11YaTaSa.PN}, in the circumstellar envelopes of evolved stars such as Asymptotic Giant Branch (AGB) stars and
red supergiants like VY Canis Majoris \citep{08MiHaTe.PN, Ziurys07, 18ZiScBe.PN}, and in cold molecular clouds with temperatures ranging from 10–100 K \citep{Rivilla18}. Of the more recent astrophysical developments worth mentioning is the first extragalactic detection of PN by \citet{22HaRiMa.PN} and the detection of PN in the edge of our Galaxy \citep{KoGoZi23}, further reinforcing the importance of PN as an astrophysical species. This unexpected prominence of PN over PO highlights gaps in our understanding of interstellar phosphorus chemistry and underscores the need for accurate spectroscopic
data to facilitate its detection and analysis.

The full summary of experimental works conducted has been thoroughly described previously \citep{Fontani24, jt590, jt842,87MiBeHe.PN, 07ToKlLi.PN, 19QiZhLi.PN, 19FoRiTa.PN, 22HaRiMa.PN, 23GoSoJa.PN, 08MiHaTe.PN, 16FoRiCa.PN, 13DeKaPa.PN}.

A recent extended \ai\ study was recently conducted by \citet{22LiZhSh.PN}, reporting 39 electronic states with spin-orbit couplings. This \ai\ study further confirmed the difficulty of modelling the \A\ state due to the multiple interactions with adjacent electronic states involving almost all vibrational levels. \citet{24TiFaWa.PN} more recently reported a variationally improved Hulburt-Hirschfelder potential model for the PN ground state, using it to predict different thermodynamic properties such as the molar heat capacity, entropy, enthalpy and Gibbs free energy.

Another recent study of \citet{22EcRiYe.PN} came up with a different methodology for producing PN, which should make future spectroscopic studies on the molecule easier. A study in photochemistry by \citet{24ZhWaJi.PN} suggested additional pathways for PN creation in the interstellar medium and star-forming regions compared to the earlier studies by \citet{20ToVexx.PN,20RiDrAl.PN,21MiSaBr.PN}.

Previously, an ExoMol line list for PN (YYLT) covering rovibrational transitions within its \X\ ground state was reported by \citet{jt590}. The YYLT line list was generated using an empirically refined potential energy curve (PEC) and an \ai\ dipole moment curve (DMC) of PN. The present study replaces the YYLT line list by both improving the treatment of the \X\ state and extending it to include the \A\ state using the standard ExoMol methodology \citep{jt939}. The UV absorption cross-sections of molecules are important not only for direct observations in the UV, but also for utilisation in atmospheric chemistry calculations \citep{jt946s}. The UV molecular data on PN will be relevant to HST, the Hubble Space Telescope (WFC3/UVIS instrument covering  200 -- 1000 $\mu$m) and GALEX, the Galaxy Evolution Explorer  (135 -- 280 nm), and the Habitable Worlds Observatory (HWO) future mission concept.

In this study, we present a systematic analysis of all experimental high-resolution transition data and derive empirical energy levels for \pn. These empirical energy levels are determined using the MARVEL (Measured-Active-Rotational-Vibrational-Energy-Levels) procedure   \citep{jt412,jt750} by inverting all \pn\ experimental frequencies obtained from the literature \citep{33CuHeHe.PN, 72HoTiTo.PN, 72WyGoMa.PN, 81GhVeVa.PN, 81MaLoxx.PN, 81CoPrxx.PN, 87VeGhIq.PN, 95AhHaxx.PN, 96LeMeDu.PN, 06CaClLi.PN} to produce a self-consistent set of experimentally derived energies for this molecule. We provide a unified account of all the experimentally suggested perturbations in \A\ \citep{81GhVeVa.PN,96LeMeDu.PN}, primarily caused by the \C, \D, \bp, \dd\ and \es\ electronic states. The \AX\ transition dipole moment is scaled to replicate the lifetimes measured by \citet{75MoMcSi.PN}. We refine the \A\ and \X\ PECs and coupling curves representing the \A--\X\ system starting from our previous \ai\ models \citep{jt842,23UsSeYu.PN} using the empirical energies to produce a semi-empirical line list PaiN for \pn\ and its isotopologue $^{31}$P$^{15}$N  valid over an extended temperature range.

\section{MARVEL}

The MARVEL algorithm \citep{jt412,12FuCsi.method,jt750,jt908} constructs a self-consistent set of rovibronic energies derived from experimental transition data. By using the experimental uncertainties for the initial weights in a weighted linear least-squares protocol, MARVEL forms associated energy levels and their respective uncertainty. Through multiple iterations in which the original weights are adjusted, MARVEL then forms a self-consistent set of energies by adjusting the uncertainty for each transition until they agree with the wider spectroscopic network (SN) \citep{12FuCsxx.methods}. This also allows one to remove the so-called ``bad lines'' -- transitions that are not in agreement with the rest of the SN. Here we use the MARVEL 4 algorithm of \citet{jt908}.

\begin{figure}
    \centering
    \includegraphics[width=0.5\textwidth]{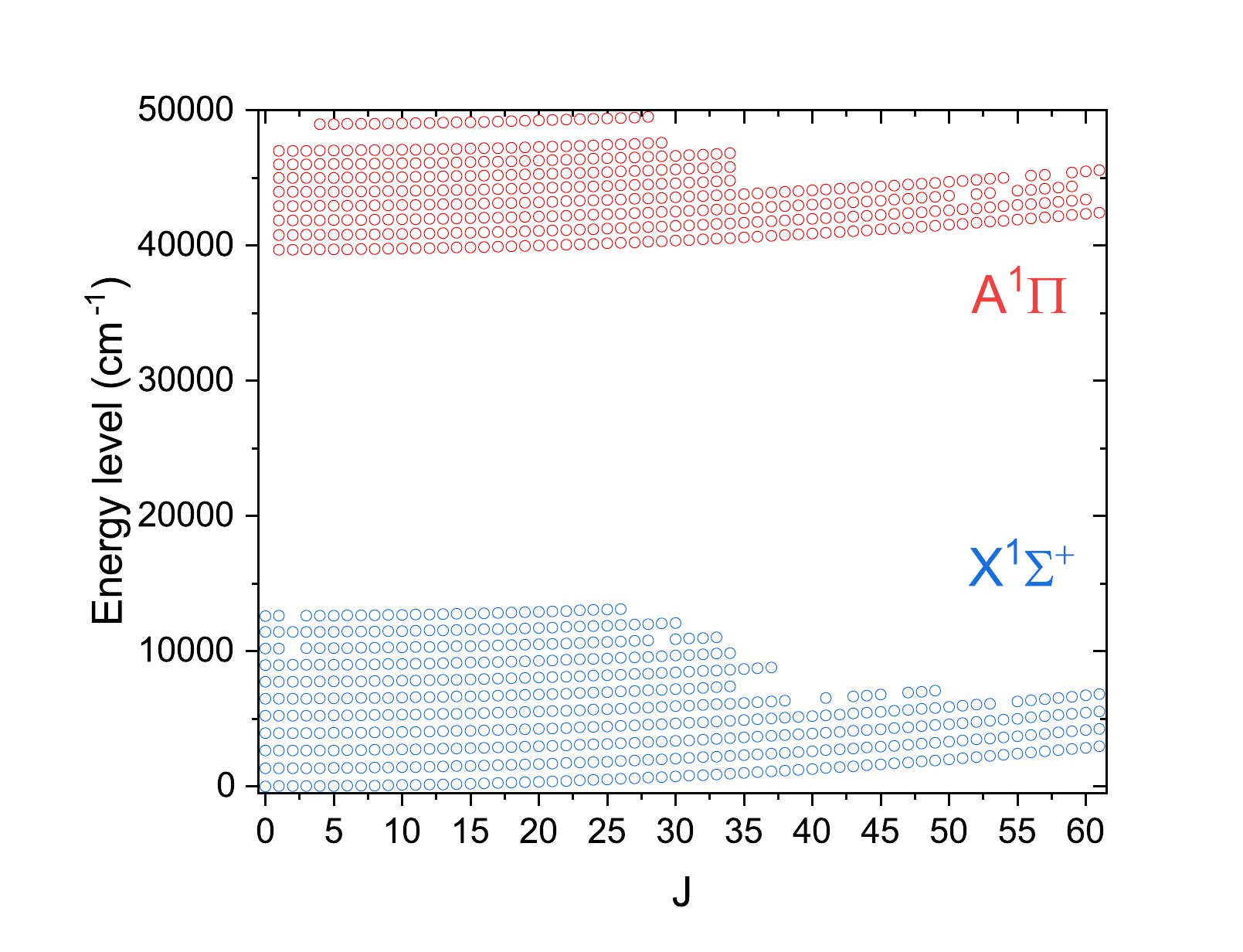}
    \caption{Empirical energy levels derived from experimental transitions using MARVEL.}
    \label{fig:marvel_energy}
\end{figure}

For \pn, all available and identifiable experimental transitions have been collected from the literature and processed using the MARVEL procedure. The transition frequencies of \pn\ extracted in this paper cover two main bands of PN involving the \X\ and \A\ electronic states, \XX, \XA, as summarised in Table \ref{t:trans}. The transition data from the \XE\ system is outside the current model, and thus, it has not been used in the MARVEL analysis. The empirical (MARVEL) energy levels of the \X\ and \A\ states of \pn\ derived are depicted in Fig.~\ref{fig:marvel_energy}.

\subsection{Experimental sources of \pn}

All the available sources of experimental transitions of \pn\ considered in this work are reviewed below, see also Table \ref{t:trans}.

\textbf{33CuHeHe} \citep{33CuHeHe.PN} is the first detection of PN in laboratory setting. In their work they reported 570 transitions of the \XA\ system for the vibrational bands (0,0),(0,1),(0,2),(1,0),(1,2) and (1,3).

\textbf{72HoTiTo} \citep{72HoTiTo.PN} reported five \X\ state hyperfine transitions $J=1 - 0$ within the $v=1$ and 0 vibrational states, which we deperturbed to two fine transition frequencies for the MARVEL analysis using the methodology outlined by \citet{jt869}.

\textbf{72WyGoMa} \citep{72WyGoMa.PN} recorded 19 \X\ state rotational transitions  within
the $v=0,1,2,3$ and 4 vibrational states using microwave spectroscopy.

\textbf{81GhVeVa} \citep{81GhVeVa.PN} reported 1680 rovibronic transitions of the \XA\ system arising from
33 $(v',v'')$ bands with $v'$ ranging from 0 to 9 and $v''$ ranging from 0 to 10.

\textbf{81MaLo} \citep{81MaLoxx.PN} reported 22 \X\ state rovibrational transitions system in the (1,0), (2,1), (3,2) and (4,3) bands.

\textbf{81CoPr} \citep{81CoPrxx.PN} reported 1382 rovibronic transitions of the \XE\ band originally with unidentified upper vibrational levels labelled simply as $v$, $v+1$, $v+2$ etc. Using the results produced by \citet{84VeGhxx.PN}, we were able to determine their reference value $v$ to be $v=4$ and assign all the other vibrational levels accordingly. The results of our assignment are included in the final experimental transition set but excluded from the MARVEL analysis  (frequencies set negative as per MARVEL procedure) as well as from the line list construction.

\textbf{87VeGhIq} \citep{87VeGhIq.PN} reported 961 rovibronic transition with unidentified bands of \pn\ $\alpha\, ^{1}\Pi$ - \X, $\beta\,^{1}\Pi$ - \X, $\delta\,^{1}\Sigma^{+}$\--\X, $\gamma\,^{1}\Sigma^{+}$ -- \X. However, because both the vibrational bands and electronic upper state assignment were undefined, this experimental data set was omitted from the MARVEL procedure at this stage but is provided as a supplementary in the MARVEL format.

\textbf{95AhHa} \citep{95AhHaxx.PN} reported 62 \X\  rovibrational transitions in the (1,0) fundamental band obtained using an FTIR (Fourier Transform Infrared) spectrometer.

\textbf{96LeMeD} \citep{96LeMeDu.PN} reported 1680 rovibronic transitions of the \XA\ band system in the (0,0), (0,1), (0,2), (1,0), (1,2), (1,3), (2,1), (2,0), (2,3), (2,4), (3,2) and (3,1) bands.

\textbf{06CaClPu} \citep{06CaClLi.PN}: reported 13 \X\ state transitions, including 12 hyperfine-resolved ones, which we deperturbed to 3 rotational transitions in the $v=0$ state for MARVEL analysis using the methodology outlined of \citep{jt869}.

\begin{table}
    \centering
    \caption{Breakdown of the assigned transitions by electronic bands for the sources used in this MARVEL study. {A} and {V}  are the numbers of the available and validated transitions, respectively. The minimum, maximum and mean uncertainties (Unc.) obtained using the MARVEL 4.0 procedure are given in \cm.}
    \label{t:trans}
    \resizebox{\textwidth}{!}{
        \begin{tabular}{lllccccc}
        \toprule
         Electronic Band & Vibrational Bands & $J$ Range & range \cm\ & A/V & Min Unc., \cm & Max Unc., \cm& Mean Unc., \cm   \\
        \midrule
        \textbf{33CuHeHe}\\
        \A--\X\ &(0,0),(0,1),(0,2), &7--65 & 36681.4 - 40758.0 & 570/199 & 0.20 & 0.48 & 0.21  \\
        &(1,0),(1,2),(1,3)\\
        \textbf{72HoTiTo}\\
        \X--\X\ & (0,0),(1,0) &0--1 & 1.556319159 - 1.567427720 & 2/2 & 4.59 $\times 10^{-9}$ & 2.88$\times 10^{-8}$ & 1.78$\times 10^{-8}$  \\
        \textbf{72WyMaGo}\\
        \X--\X\ &(0,0), (1,1), (2,2), & 2--8 & 3.112628 - 12.537217 & 19/19 & 6.67$\times 10^{-6}$ & 6.67$\times 10^{-6}$ & 6.67$\times 10^{-6}$  \\
        & (3,3), (4,4)  \\
        \textbf{81MaLo}\\
        \X--\X\ &(1,0), (2,1), (3,2),& 2--52 & 1217.216 - 1318.591 & 22/22 & 3.00$\times 10^{-4}$ & 3.10$\times 10^{-3}$ & 1.38$\times 10^{-3}$  \\
        &(4,3)  \\
        \textbf{81CoPr}\\
        \E--\X\ &(4,3),(5,2),(6,1) & 0--62 & 52355.10 - 61787.57 & 1382/0 & 5.00$\times 10^{-2}$ & 5.00$\times 10^{-2}$ & 5.00$\times 10^{-2}$  \\
        &(6,2),(6,5),(7,1),\\
        &(8,0),(8,1),(8,4),\\
        &(9,0),(9,3),(10,0),\\
        &(10,2),(11,0)\\
        \textbf{81GhVeVa}\\
        \A--\X\ &(0,0),(0,1),(0,2) & 0--39 & 33334.96 - 43081.72 & 2188/1586 & 5.00$\times 10^{-2}$ & 2.76$\times 10^{-1}$ & 5.18$\times 10^{-2}$  \\
        &(1,0),(1,2),(1,3),\\
        &(1,4),(2,0),(2,3),\\
        &(2,4),(2,5),(3,1),\\
        &(3,2),(3,3),(3,4),\\
        &(3,5),(3,6),(4,1),\\
        &(4,2),(4,5),(4,6),\\
        &(4,7),(4,8),(5,2),\\
        &(5,7),(5,8),(5,9),\\
        &(6,3),(6,9),(6,10),\\
        &(7,3),(7,10),(9,5)\\
        \textbf{95AhHa}\\
        \X--\X\ & (1,0) & 0--33 & 1273.296 - 1368.494 & 62/61 & 2.00$\times 10^{-3}$ & 2.00$\times 10^{-3}$ & 2.00$\times 10^{-3}$ \\
        \textbf{96LeMeDu}\\
        \A--\X\ &  (0,0),(0,1),(0,2),& 1--69  & 36459.57 - 41856.93 & 1680/959 & 5.00$\times 10^{-2}$ & 3.12$\times 10^{-1}$ & 5.14$\times 10^{-2}$  \\
        &(1,0),(1,2),(1,3),\\
        &(2,1), (2,0), (2,3),\\
        &(2,4),(3,2),(3,1)\\
        \textbf{06CaClPu}\\
        \X--\X\ &(0,0), (1,1), (2,2),&0--52 & 1.5674270 - 10.9705230 & 16/16 & 2.00$\times 10^{-7}$ & 1.33$\times 10^{-6}$ & 9.70$\times 10^{-7}$  \\
        &(3,3), (4,4), (1,0),\\
        &(2,1), (3,2), (4-3) \\
        \bottomrule
        \end{tabular}
    }
\end{table}

\subsection{\A\ perturbations}
\label{s:perturbations}

As can be seen in previous \ai\ works by \citet{14AbSaMa.PN,19QiZhLi.PN,22LiZhSh.PN}, the \A\ state is surrounded by multiple electronic states within which it interacts with, including \D, \es\  and \dd, see Fig.~\ref{fig:currentPECs}. The spin-orbit couplings with these dark states are of the order 20--30~\cm, see \citet{jt842}.

The experimental evidence given by \citet{81GhVeVa.PN} and \citet{96LeMeDu.PN} points to multiple crossings for $v=0,1,2,3$ and $7$. It has also been suggested that \A\ is perturbed by the \E\ state at $v=14$ by \citet{22LiZhSh.PN}, but there is no other experimental evidence in the required region to confirm this. Below we provide a detailed description of the state crossings and resulting perturbations for each known vibrational state of \A\ affected.

\begin{figure}
    \centering
    \includegraphics[width=0.45\textwidth]{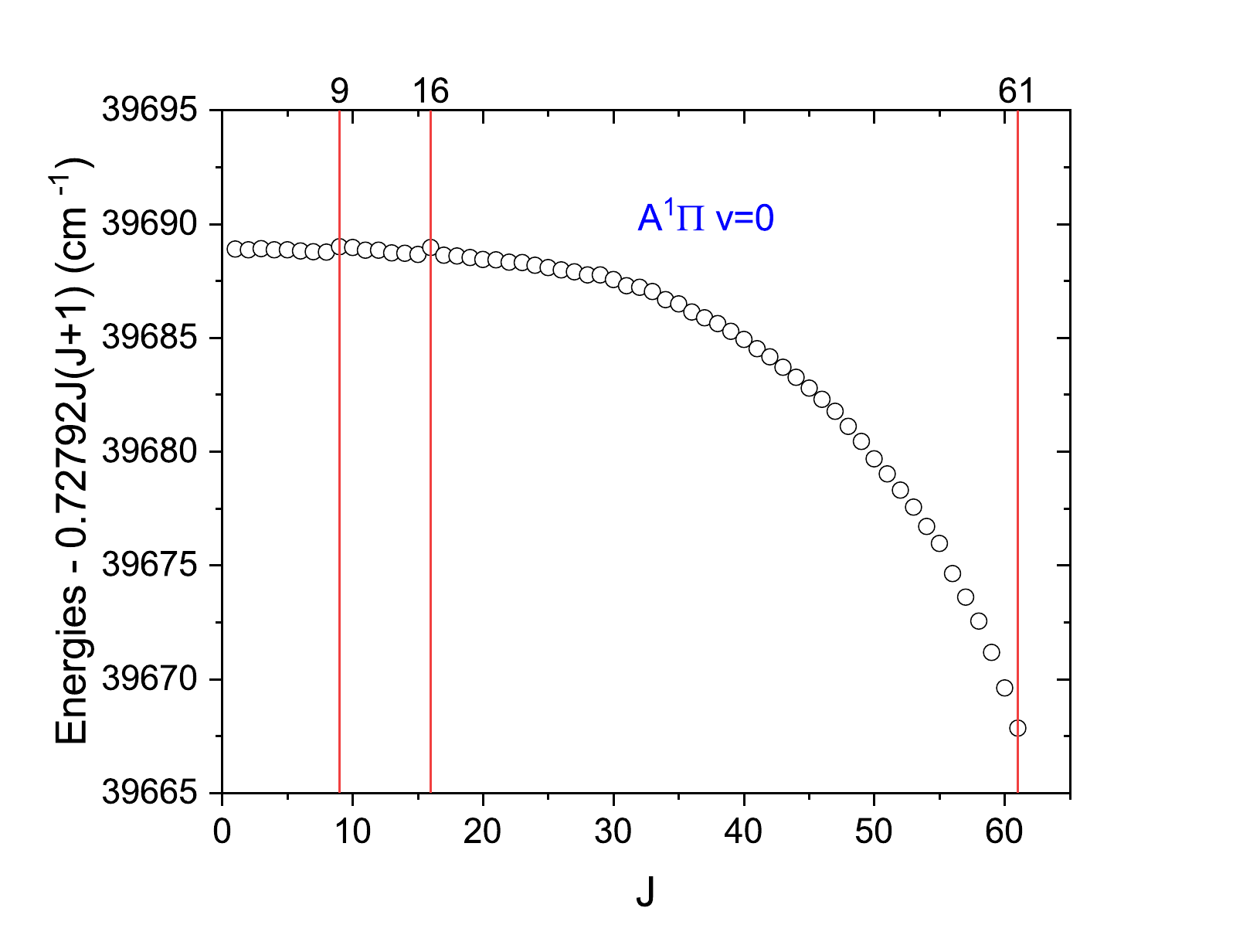}
    \includegraphics[width=0.45\textwidth]{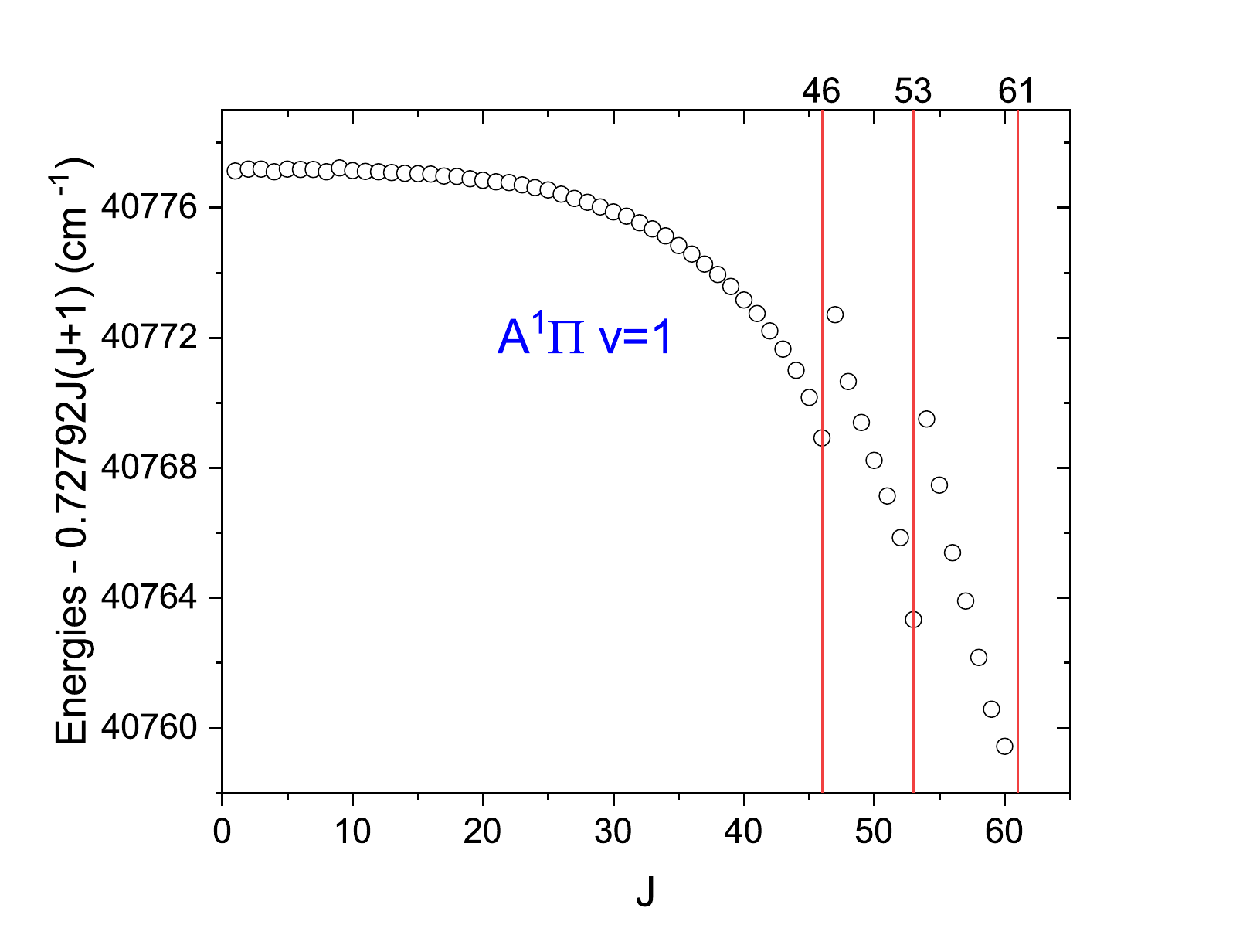}\\
    \includegraphics[width=0.45\textwidth]{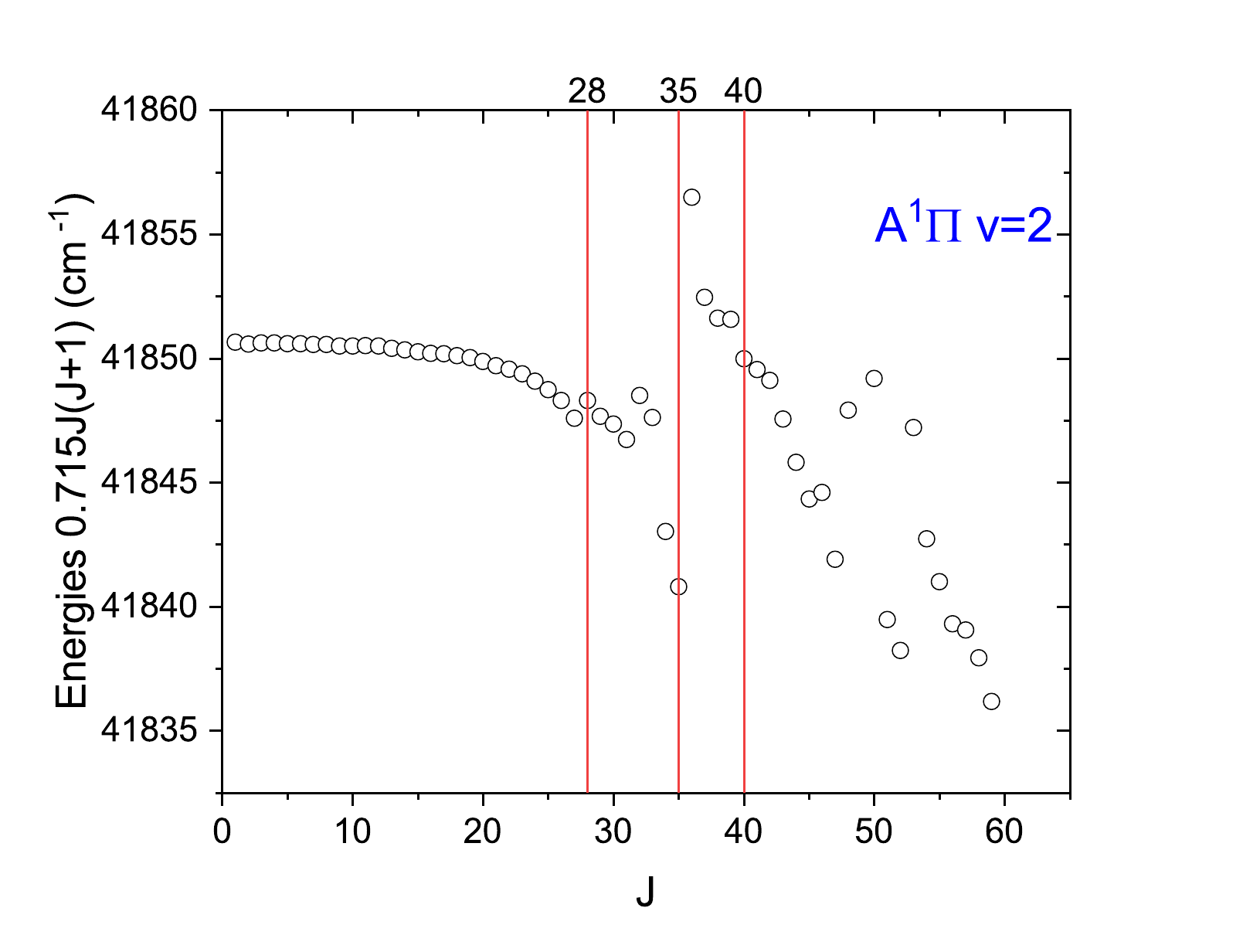}
    \includegraphics[width=0.45\textwidth]{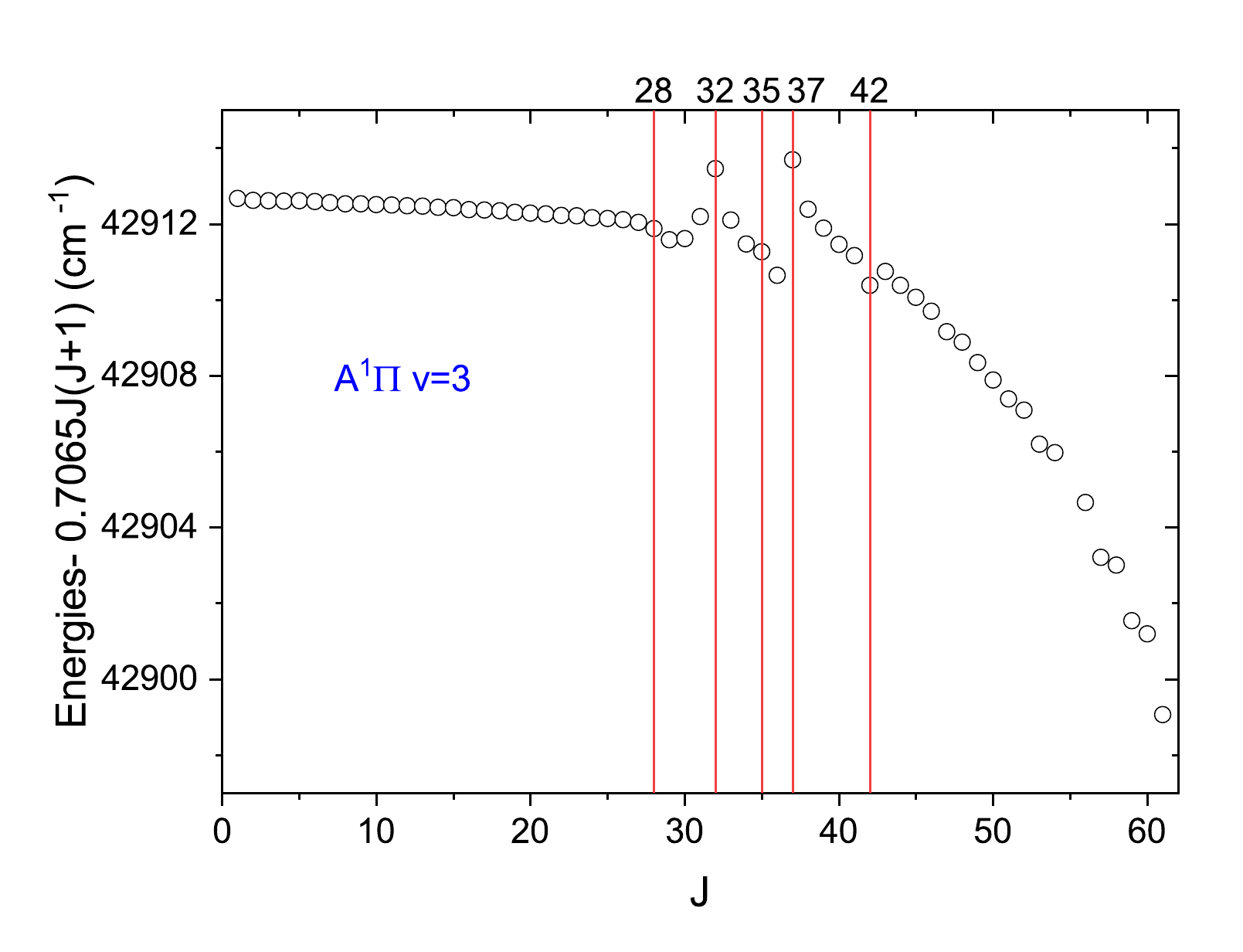}\\
    \includegraphics[width=0.45\textwidth]{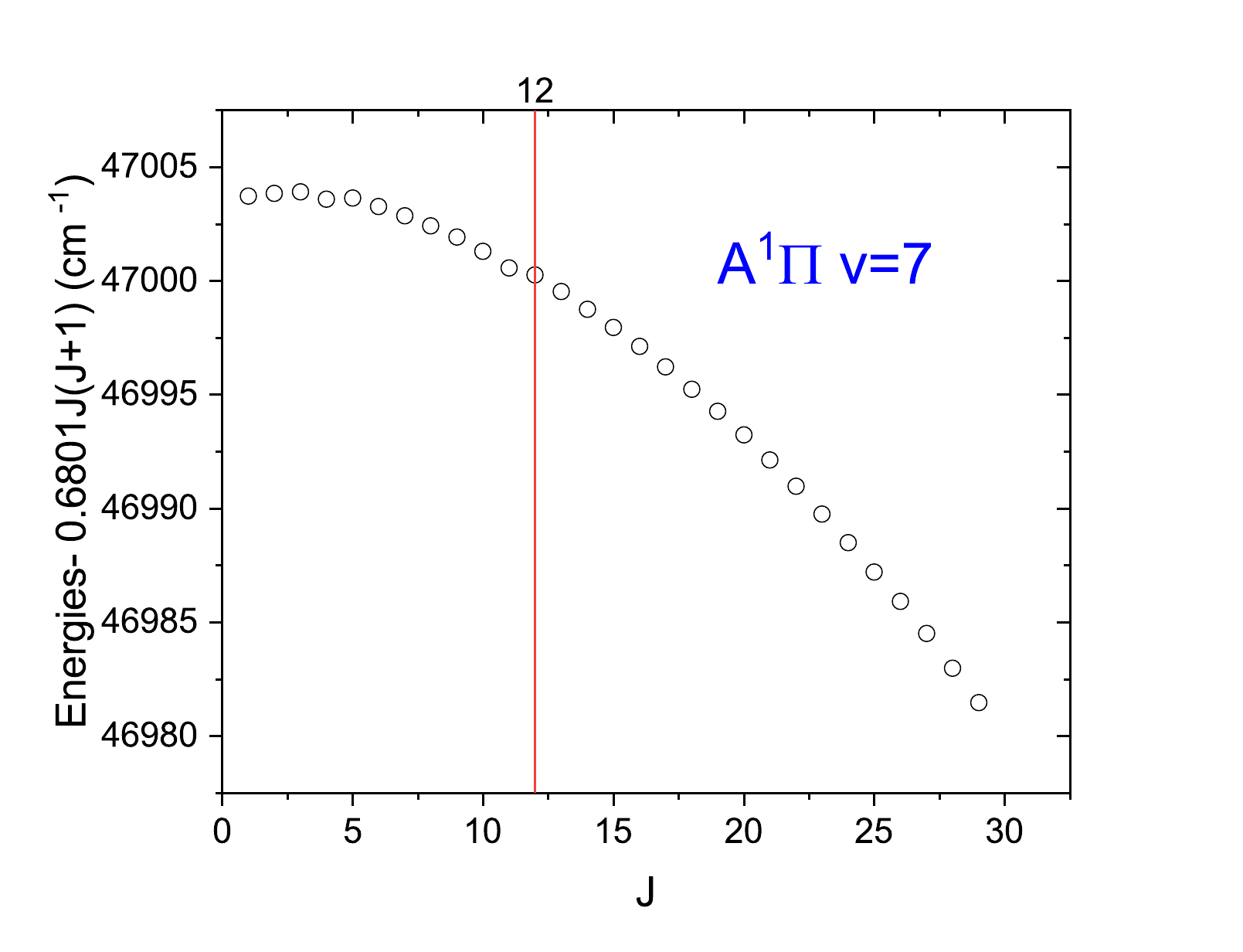}
    \caption{Reduced empirical energy term values for the $v=0,1,2,3,7$ of the \A\ state of PN in the rotational decomposition. The rotational $B$ constants used for each $v$ were taken from \citet{81GhVeVa.PN}. The lines indicate  $J$ values corresponding to state crossings as suggested by \citet{81GhVeVa.PN} and \citet{96LeMeDu.PN}. }
    \label{fig:rotdecomp}
\end{figure}

The \A\ $v=0$ state is believed to be perturbed by \D, \es\ \citep{96LeMeDu.PN} and \dd\  \citep{81GhVeVa.PN}. There is general agreement on the crossing at $J=16$, but \citet{96LeMeDu.PN} could not verify the suggested crossing by \citet{81GhVeVa.PN} at $J=9$ by \dd\ and the work of \citet{81GhVeVa.PN} also does not extend to $J=61$ to confirm the \es\ crossing suggested by \citet{96LeMeDu.PN}. Both the $J=9$ and $J=16$ suspected crossings are confirmed within our MARVEL data by means of the rotational decomposition of energy levels, as can be seen in Fig.~\ref{fig:rotdecomp}.

The \A\ $v=1$ state is perturbed by the \es\ state at $J=46$ and 53 as predicted by \citet{81GhVeVa.PN} and at $J=46,53$, and 61 suggested by \citet{96LeMeDu.PN}. These perturbations are  confirmed in the rotational decomposition of our reduced MARVEL energies for $v=1$ shown in Fig.~\ref{fig:rotdecomp}.

The \A\ $v=2$ state is heavily perturbed by the \es\  state at $J=28,35$ and 40  as predicted both by \citet{81GhVeVa.PN} and \citet{96LeMeDu.PN}; however, the impact of the crossing with the \es\  state is much more profound than in the $v=1$ state. The crossing points are confirmed by us and shown in Fig.~\ref{fig:rotdecomp}.

The \A\ $v=3$ state is believed to be perturbed by \bp\ or \D\ at $J=28$ according to \citet{81GhVeVa.PN}, whereas \citet{96LeMeDu.PN} believes that \dd\ and \C\ are the main perturbing electronic states at $J=32,37,35$ and 42. Our MARVEL data agrees on the crossings suggested by \citet{96LeMeDu.PN} even though it is hard to distinguish separate perturbations at $J=35$ and $J=37$. Our MARVEL $v=3$ energy is perturbed at $J=28$, it is hard to verify if the crossing is by  \bp\ or \D\ as suggested by \citet{81GhVeVa.PN}. All the suggested crossings are illustrated in the reduced energy diagram in Fig.~\ref{fig:rotdecomp}.

Both \citet{81GhVeVa.PN} and \citet{96LeMeDu.PN} suggest that the \A\ $v=7$ state is crossed by the \D\ at $J=11$. The resulting perturbation is much weaker than for the lower lying \A\ $v$ states but can still be seen in Fig.~\ref{fig:rotdecomp}.


The multiple crossings involving the \A\ state,  which are summarised in Table \ref{tab:A_experimental_perturbations}, make it hard to model the spectra.
An advantage of the MARVEL approach is that it does not depend on a model and is, therefore, able
to capture the energies of the perturbed levels correctly; something that is very hard to do
with effective Hamiltonians. Our spectroscopic model also struggles to reproduce the many
perturbations as there is little information on the perturbing states. However, observed transitions involving perturbed levels with $J>30$ for $v=1,2$ and 3 did not always yield a consistent spectroscopic network (SN), which led us to exclude multiple transitions with higher $J$ as can be seen in Fig.~\ref{fig:marvel_energy}. Ideally, we would be able to assign the affected transitions to the dark states, stealing intensity from the \A\ state, and fully model the interactions; however, because of the ambiguous experimental assignment of vibrational levels of the interacting dark states, we elected to not include these in our MARVEL SN. This results in us sacrificing the quality of the calculated energies around the crossings within our spectroscopic model, but also providing a better fit overall to the remaining energies.

\begin{table}
    \centering
    \caption{Summary of positions of the crossings $J_{\rm c}$ of the vibronic states $v$, \A\  caused by perturbations by other electronic states as suggested by different experimental studies.}
    \begin{tabular}{c|c|c|c}
    \toprule
    \bf{\A\ }& & &\\
        $v$ & $J_{\rm c}$  & Suggested perturbing state(s) &  Source\\
    \midrule
        0 &  8/9 & \dd\ & \bf{81GhVeVa} \\
        0 & 15/16 & \D\ & \bf{96LeMeDu} \\
        0 &  16/17 & \dd\ & \bf{81GhVeVa} \\
        0 & 61/62 & \es\ & \bf{96LeMeDu} \\
    \midrule
        1 & 45/46 & \es\ & \bf{81GhVeVa} \\
        1 & 46/47 & \es\ & \bf{96LeMeDu} \\
        1 & 52/53 & \es\ & \bf{81GhVeVa} \\
        1 & 53/54 & \es\ & \bf{96LeMeDu} \\
        1 & 60/61 & \es\ & \bf{96LeMeDu} \\

     \midrule
        2 & 27/28 & \es\ & \bf{96LeMeDu} \\
        2 & 27/28 & \es\ & \bf{81GhVeVa} \\
        2 & 35/36 & \es\ & \bf{96LeMeDu} \\
        2 & 35/36 & \es\ & \bf{81GhVeVa} \\
        2 & 40/41 & \es\ & \bf{96LeMeDu} \\
     \midrule
        3 & 28/29 & \bp\, \D\ & \bf{81GhVeVa} \\
        3 & 31/32 & \dd\ & \bf{96LeMeDu} \\
        3 & 36/37 & \dd\ & \bf{96LeMeDu} \\
        3 & 34/35 & \C\ & \bf{96LeMeDu} \\
        3 & 42/43 & \dd\ & \bf{96LeMeDu} \\
     \midrule
        7 & 11/12 & \D\ & \bf{96LeMeDu, 81GhVeVa} \\
    \bottomrule
    \end{tabular}
    \label{tab:A_experimental_perturbations}
\end{table}

\section{Spectroscopic Model and Refinement}

For our spectroscopic model and its subsequent refinement, we use the variational diatomic nuclear-motion code \Duo\footnote{Freely available at \href{https://github.com/exomol/Duo}{github.com/exomol/Duo}} \citep{Duo} to solve the coupled system of  Schr\"{o}dinger equations for the \X\ and \A\ systems of PN. In these calculations, we used the Sinc DVR method for the vibrational degree of freedom on a grid of 701 points ranging from 1.0 to 5.0~\AA.

Our spectroscopic model selection for the description of the \AX\ system is fully uncoupled from the other electronic states of PN. The corresponding  PECs and couplings are shown in Figures \ref{fig:currentPECs} and \ref{fig:couplings}. In Fig.~\ref{fig:currentPECs}, we also show \ai\ curves corresponding to other electronic states from this region from our previous work \citep{jt842}, not considered in the current model: \as,  \bp,  \dd, \es,   \C, \D\ and \E. Except for \C\ and \E, these are all ``dark'' states with respect to the transitions from/to the ground electronic state \X. Indeed, \as, \es,  \bp\ and \dd\ have different multiplicities and are therefore dipole forbidden (before the coupling), while the \D\ state is also dipole forbidden due to the selection rule
$$
\Delta \Lambda= 0, \pm 1.
$$
Here $\Lambda$ is the projection of the electronic angular momentum on the molecular axis and $\Lambda=0, \pm 1, \pm 2$ for $\Sigma$, $\Pi$ and $\Delta$, respectively. Although the \C\ and \E\ states are not dipole forbidden for \X, they were deemed to be unimportant for the \AX\ system and hence also excluded. The \C--\X\ band is weak because of the small Franck-Condon factor (see also \citet{jt842}, while the \EX\ band is off by about 20000~\cm. The main reasons for not including these states are (i) insufficient quality of the \ai\ data, (ii) a large number of local perturbations within the \A\ region (see Section~\ref{s:perturbations}) and (iii) a severe lack of the experimental data required for proper description of the underlying PECs and other couplings. Therefore, in this work, for the spectroscopic description of the  \AX\ system, we opted for a complete exclusion of these interactions instead of having them all in the wrong places.

\subsection{PECs}

As a starting point for this model, the results from the two previous studies \citep{jt842, 23UsSeYu.PN} were selected for refinement using energy levels from MARVEL. For the \X\ state, we used an empirical potential energy curve (PEC) of PN taken from  \citet{23UsSeYu.PN} where it was represented by the Extended Hulburt-Hirschfelder (EHH) potential as given by
\begin{equation}
\label{e:EHH}
    V_{\textrm{EHH}}(r)=D_\textrm{e}\left[\left(1-e^{-q}\right)^2 + cq^3\left(1+\sum_{i=1}^3 b_iq^i\right)e^{-2q}\right],
\end{equation}
with $D_{\rm e}$ as a dissociation energy and $q=\alpha \left(r-r_\textrm{e}\right)$. $D_{\rm e}$ was kept at the value of 51297 \cm\ calculated by \citet{23UsSeYu.PN}.

For the \A\ state,  the initial PEC, calculated using internally contracted multireference configuration interaction with a Davidson correction (icMRCI+Q) with the aug-cc-pV5Z-DK basis set, was taken from our previous work \citep{jt842} and represented analytically by an  Extended Morse Potential (EMO) \citep{EMO}, which is described by:
\begin{equation}
\label{e:EMO}
V(r)=V_{\mathrm{e}}+(A_{\mathrm{e}}-V_{\rm e})\left(1-\exp \left[-\left(\sum_{i=0}^N a_i \xi_p(r)^i\right)\left(r-r_{\mathrm{e}}\right)\right]\right)^2,
\end{equation}
where $A_{\mathrm{e}}$ is the dissociation asymptote of \A, $V_{\rm e}$ is the potential minimum, $a_i$ are expansion coefficients, $r_{\rm e}$ is the equilibrium bond length of the electronic state, and $\xi_p(r)$ is a \v{S}urkus variable \citep{84SuRaBo.method} given by:
\begin{equation}
\label{e:surkus}
    \xi_p(r)=\frac{r^p-r_{\mathrm{e}}^p}{r^p+r_{\mathrm{e}}^p}.
\end{equation}
The dissociation asymptote, $A_{\rm e}$, coming from the dissociation channel P(${}^{2}$D)+N(${}^{2}$D) was set to 82~500 \cm\ as calculated from the atomic energies using the NIST atomic spectra database \citep{NIST} rounded to 3 significant figures, while  $V_{\rm e}$, $r_{\rm e}$ and $a_i$ are treated as adjustable parameters.

\subsection{Couplings}

To construct a high accuracy model while also restricting the model to the two states for which we have extensive empirical data, we added two couplings. Firstly we included an electronic angular momentum coupling curve (EAMC) $\bra{X^1\Sigma^+}$ $L_x$ $\ket{A^1\Pi}$, which we took from our previous work \citep{jt842}. The \ai\ EAMC  was given an analytical shape via the morphing functionality in \Duo, by using a polynomial decay form given as:
\begin{equation}
\label{e:F(R)}
F(r)=\sum_{k=0}^N B_k z^k\left(1-\xi_p\right)+\xi_p B_{\infty},
\end{equation}
with
\begin{equation}
z=\left(r-r_{\text {ref }}\right) e^{-\beta_2\left(r-r_{\text {ref }}\right)^2-\beta_4\left(r-r_{\text {ref }}\right)^4},
\end{equation}
where $r_{\rm ref}$ is a reference coordinate for the expansion, in our case set to the minimum of the ground state potential, $\beta_2$ and $\beta_4$ are damping factors, $B_k$ are expansion coefficients with $B_{\rm \infty}$ typically set to 1 and $\xi$ is a \v{S}urkus variable as in Eq.~(\ref{e:surkus}).

A $\Lambda$-doubling empirical curve was also included for the \A\ state, using the ``lambda-q'' functionality in \Duo. It was also given an analytical form of Eq.~\eqref{e:F(R)}  with a single expansion term.

\subsection{Refinement}

\begin{figure}
    \centering

    \includegraphics[width=0.7\textwidth]{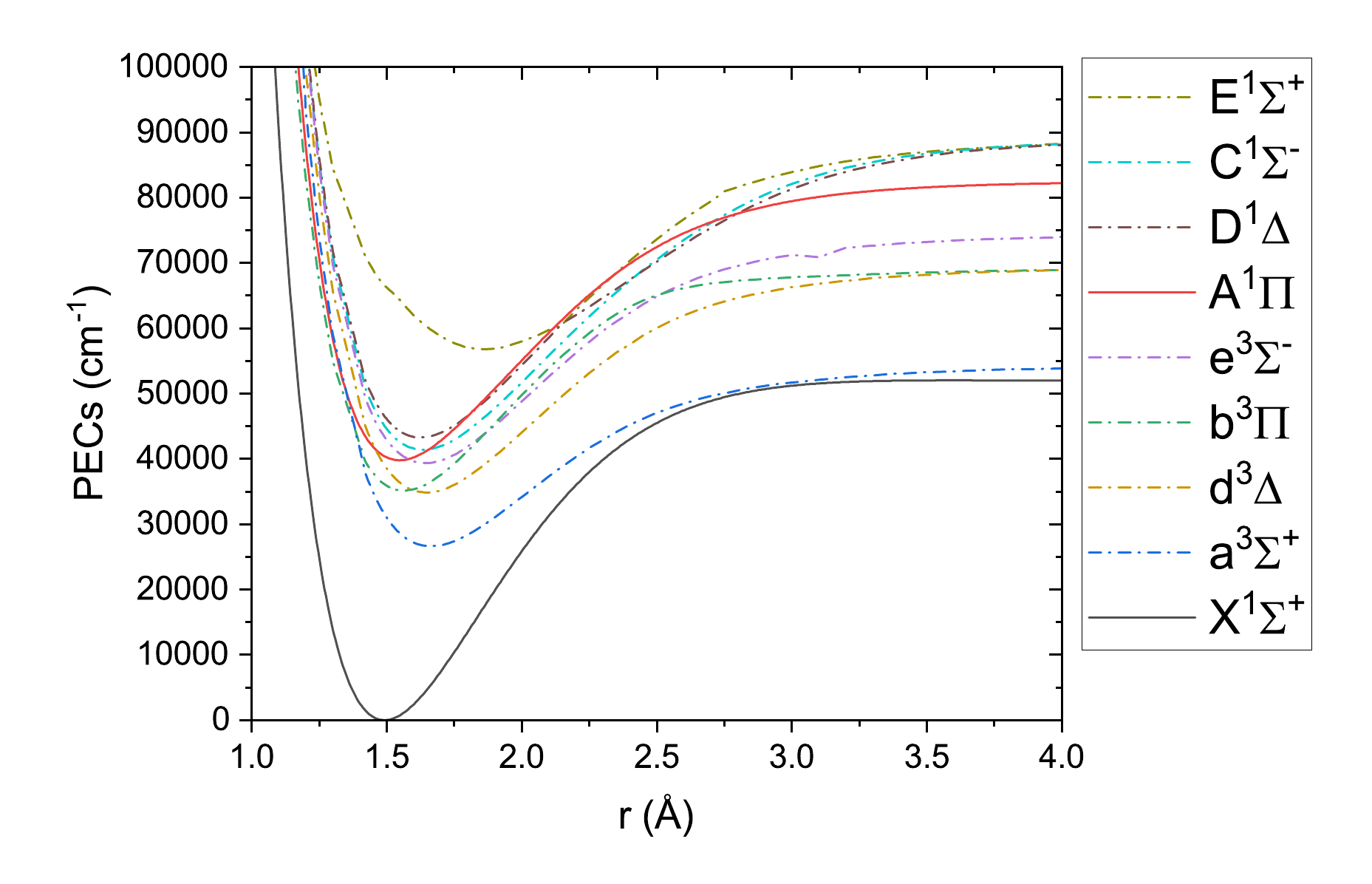}

    \caption{Empirical PECs of PN used as part of our new spectroscopic model (solid lines) and \ai\ PECs of other electronic states of PN from our previous work \citep{jt842}. \green{MS} Table 1 of the previous work also contains \ai\ spectroscopic constants that could be helpful for the reader.
    }
    \label{fig:currentPECs}
\end{figure}

\begin{figure}
    \centering
    \includegraphics[width=0.45\textwidth]{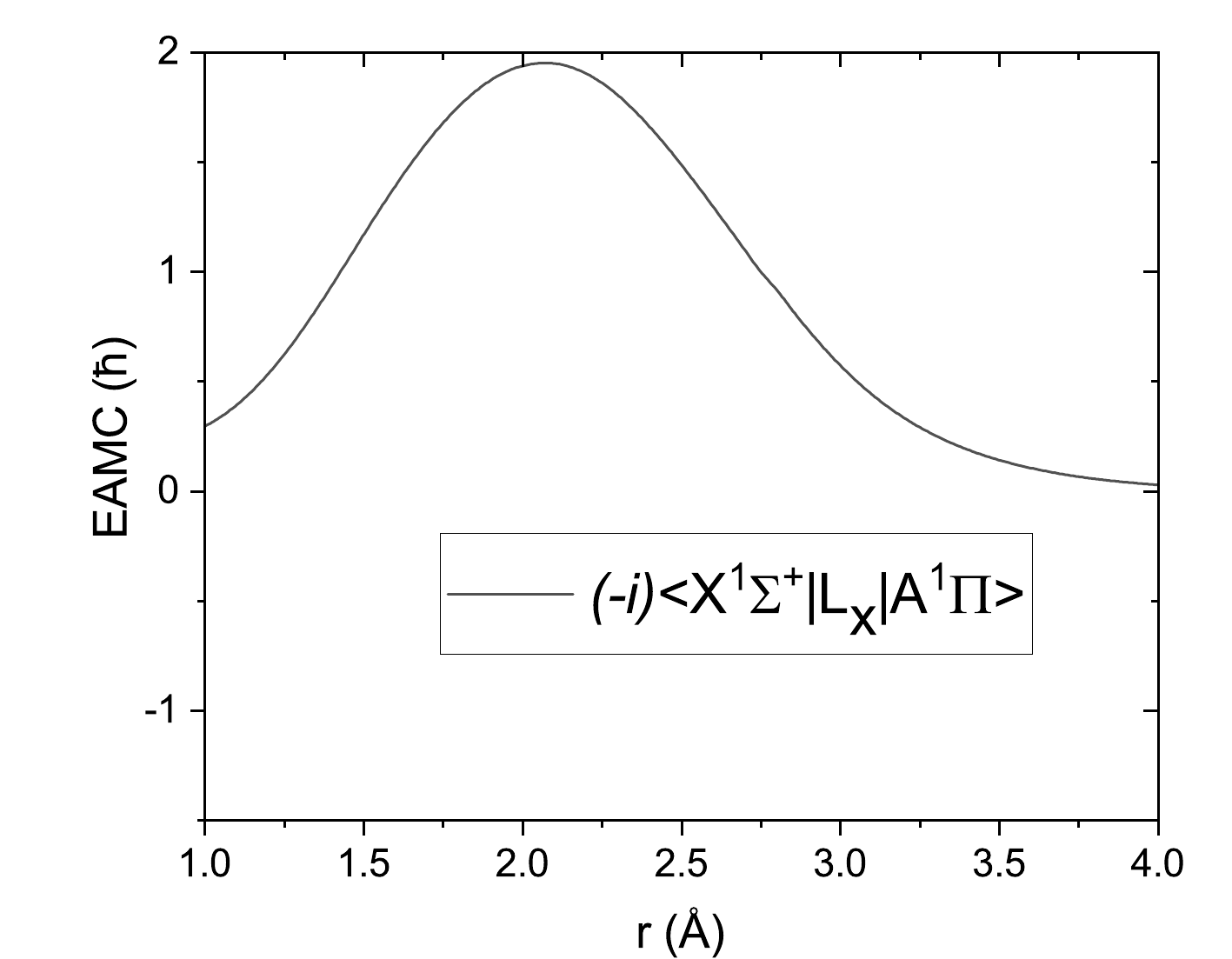}
    \includegraphics[width=0.45\textwidth]{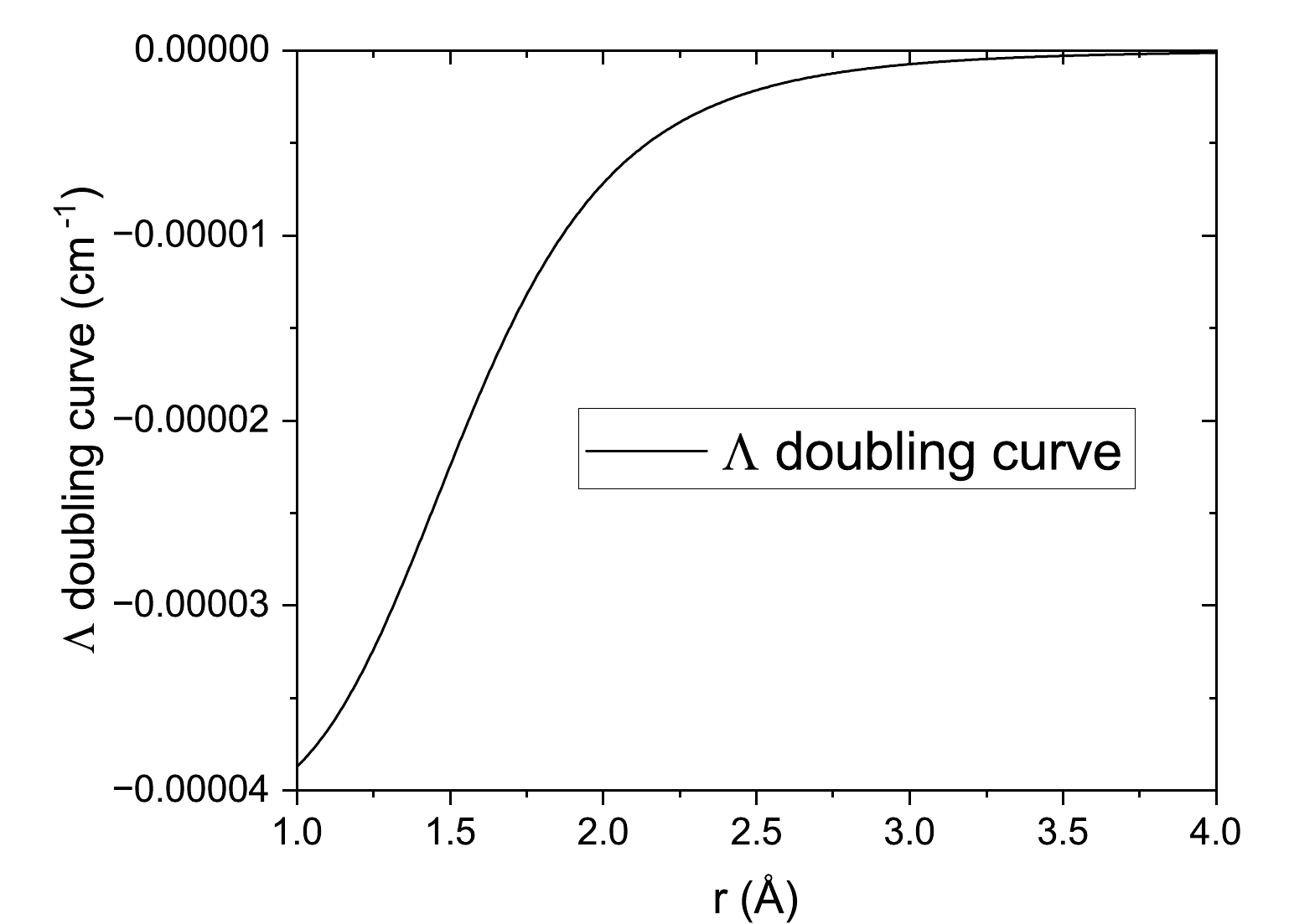}
    \caption{Fitted Electronic Angular Momenta and $\Lambda$-Doubling coupling curves of PN used in our spectroscopic model.}
    \label{fig:couplings}
\end{figure}

We refined our spectroscopic model of PN represented by the PECs and coupling curves to the MARVEL energies. Overall, 12 parameters were fitted across 4 functions to 1224 energy levels from MARVEL with the angular momentum $J$ ranging from 0 to 61. These parameters as part of the overall fit are given in the supplementary material as part of the final \Duo\ input file used for the calculations. The fitted PECs and couplings can be seen in Figures \ref{fig:currentPECs} and \ref{fig:couplings}. The PEC of the \X\ from \citet{jt590} was also refined because of the expanded range of available energy level data to fit to and to account for the electronic angular momenta coupling with the \A\ state.

\begin{figure}
    \centering
    \includegraphics[width=0.45\textwidth]{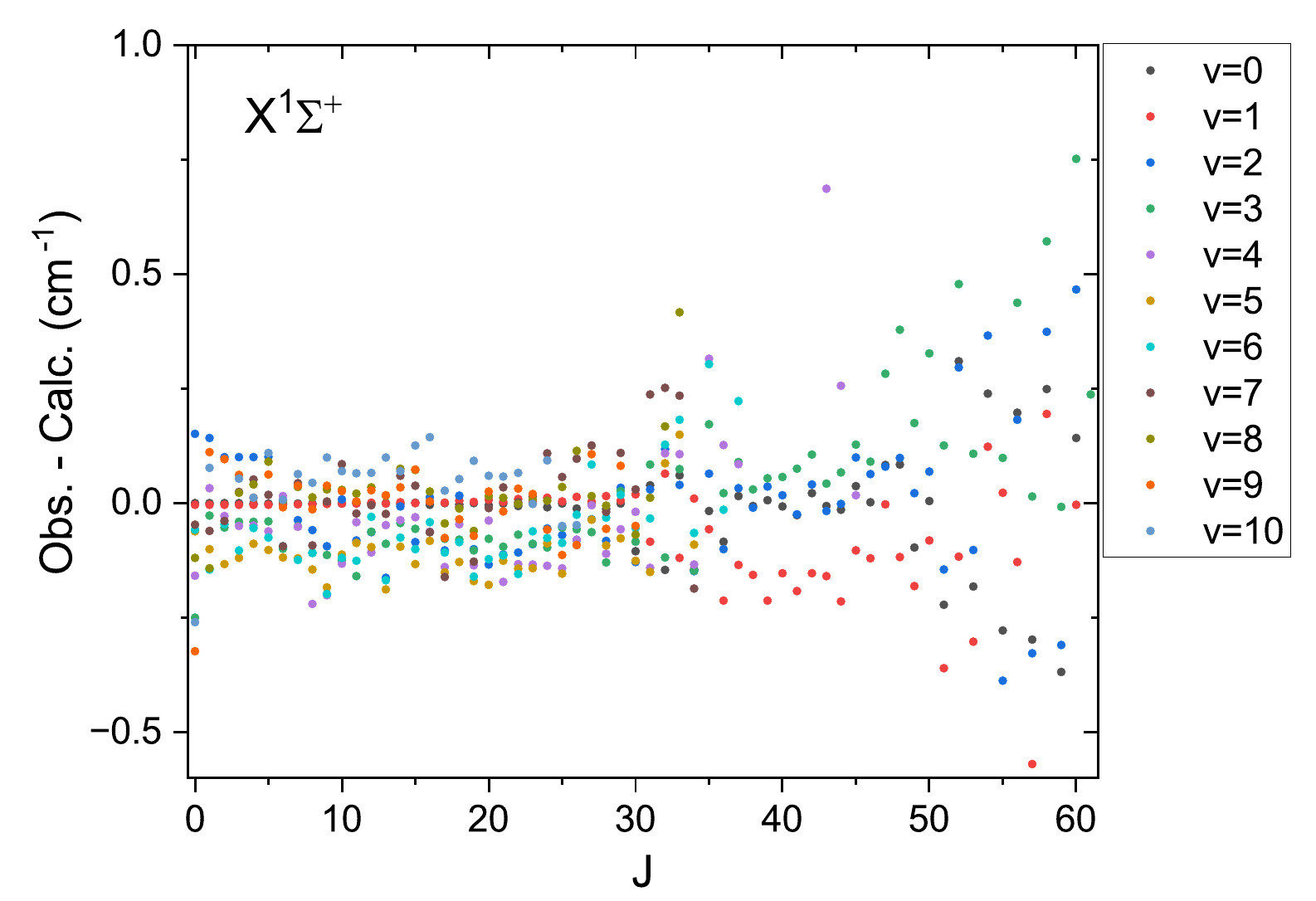}
    \includegraphics[width=0.45\textwidth]{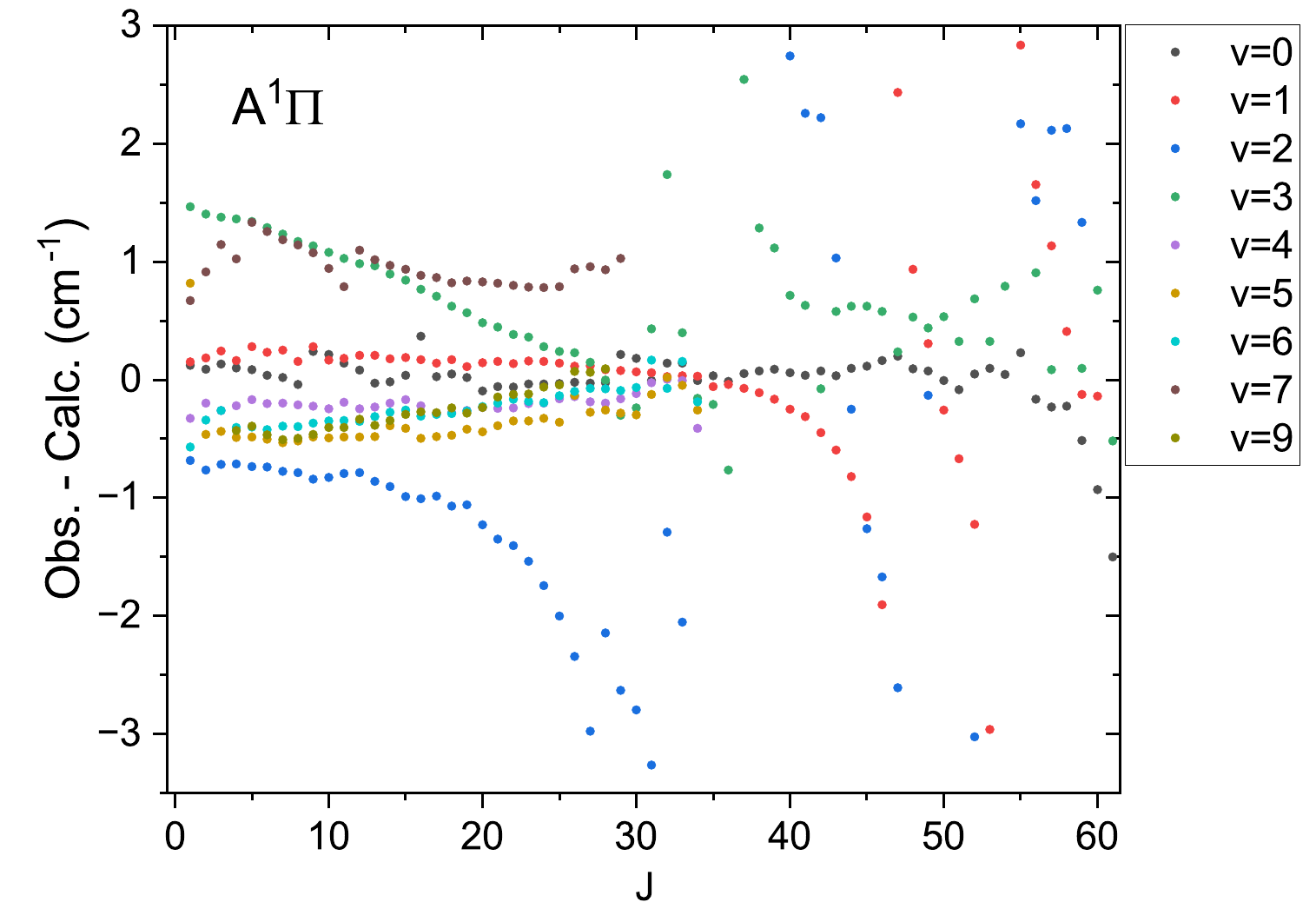}

    \caption{Visual representation of the MARVEL residuals (Obs. - Calc.) for the \A\ and \X\ state levels. }
    \label{fig:marvelresiduals}
\end{figure}

The root-mean-square (rms) error of the overall fit is 0.971 \cm, where the rms error of the \X\ state and \A\ state energies are 0.26 and 1.24 \cm, respectively. In Table \ref{tab:rms_x}, we provide rms errors of the calculated energy term values of the \X\ state, comparing the current model and energies from the previous YYLT line list \citep{jt590}. While there is a slight increase in the rms for the $v=0,1$, it can be seen that the $v=2 \dots\ 10 $ are better fitted within our model. It is worth noting that in the original paper by \citet{jt590}, the model was fitted to data based on 147 lines ($v\le 4$) and effective vibrational term values ($v\le 11$, $J=0$), whereas in our current work we have expanded that to 2,838 active lines as described in the MARVEL section of this paper with the overall more extended $J$ and $v$ coverage.  We also provide an overview of the rms error of the \A\ state of our model in Table \ref{tab:rms_a}. The majority of the residuals driving the rms of the fit are the perturbed $v$ states of \A\ as can be seen in Fig.~\ref{fig:marvelresiduals} (see also the discussion of the perturbations of \A\ in Section~\ref{s:perturbations}). 
The sources of perturbations are believed to be dark electronic states, namely \es,  \bp, \dd, \D\  and \C\ \citep{81GhVeVa.PN,96LeMeDu.PN}, see Fig.~\ref{fig:currentPECs} and also discussion in Section \ref{s:perturbations}, and are especially strong around the predicted intersections listed in Table \ref{tab:A_experimental_perturbations}.

\begin{table}
    \centering
    \caption{Root mean square errors (RMSE, \cm)  of the calculated energy term values of the \X\ state as compared to the experimentally derived (MARVEL) values ($v=0\ldots 10$) for the rovibrational \Duo\ values from the current model (\name) and the YYLT values \citep{jt590}.}
    \begin{tabular}{lrr}
    \toprule
     & \multicolumn{2}{c}{RMSE} \\
    \toprule
    $v$ & \name & YYLT \\
    \midrule
    0 & 0.0038 & 0.0029 \\
    1 & 0.0057 & 0.0053 \\
    2 & 0.0888 & 0.0994 \\
    3 & 0.0925 & 0.1257 \\
    4 & 0.1053 & 0.1703 \\
    5 & 0.1233 & 0.2266 \\
    6 & 0.1033 & 0.2196 \\
    7 & 0.0694 & 0.1358 \\
    8 & 0.0543 & 0.0802 \\
    9 & 0.0828 & 0.0934 \\
   10 & 0.0892 & 0.2938 \\
    \bottomrule
    \end{tabular}
    \label{tab:rms_x}
\end{table}

\begin{table}
    \centering
    \caption{Root mean square errors (RMSE, \cm)  of the calculated energy term values of the \A\ state from the current model as compared to the experimentally derived (MARVEL) values ($v=0\ldots7,9$).} 
    \begin{tabular}{rr}
    \toprule
    $v$ & RMSE \\
    \midrule
    0 & 0.276 \\
    1 & 1.495 \\
    2 & 2.666 \\
    3 & 0.871 \\
    4 & 0.214 \\
    5 & 0.502 \\
    6 & 0.295 \\
    7 & 0.973 \\
    9 & 0.313 \\
    \bottomrule
    \end{tabular}
    \label{tab:rms_a}
\end{table}

\subsection{(Transition) dipole moment curves}

The \X\ state permanent DMC of  \citet{23UsSeYu.PN} was used. It was  represented using the following functional form:
\begin{equation} \label{e:irreg}
    d_{\textrm{irreg}}(r;n,k) = \frac{\left[1-\exp(-\alpha r)\right]^n}
    {\sqrt{\left(r^2-a_1^2\right)^2+b_1^2} \sqrt{\left(r^2-a_2^2\right)^2+b_2^2}}\sum_{i=0}^kc_i\left(1-2e^{-\beta r}\right)^i
\end{equation}
as recommended by \citet{Medvedev22}. By using this form, we aim to reduce the numerical noise in the intensities of high overtones as well as the associated saturation at high wavenumbers, leading to the so-called ``overtone plateaus'' \citep{16MeMeSt}.

For the transition dipole moment curve (TDMC) of the \XA\ system,  we chose the ECP10MWB \ai\ dipole calculated using icMRCI+Q from our previous work \citep{jt842}, where it was shown to provide the closest lifetimes to the experiment. The \ai\ TDMC was given on a grid and then scaled through the \Duo\ ``morphing'' by a factor of 1.28 to match the experimental lifetimes as described in Section \ref{s:lifetimes}.



\section{Line list and Spectra}
\label{s:linelist}

The \name\ line lists for two isotopologues of PN, \pn\ and ${}^{31}$P${}^{15}$N,  were produced with \textsc{Duo} using the empirically refined and \ai\ curves as described above. For the main isotopologue \pn, it contains 1~333~445 transitions and 30~327 states for the  \X\ and \A\ states covering the wavenumber range up to 82~500 \cm\ (i.e. up to the dissociation of \A),  $v = 0 \ldots 63$ for \X\ and $v = 0 \ldots 74$  and $J = 0 \ldots  210$. For $^{31}$P$^{15}$N, only the atomic mass of  $^{15}$N was changed in the \textsc{Duo} calculations. We simply do not have enough experimental data for any isotopologue extrapolation procedures, described in \citet{jt948}. Further details on the number of states and transitions in each isotopologue are given in Table~\ref{t:iso}.

\begin{table}
\centering
\caption{Line list statistics for each isotopologue of PN.}
\label{t:iso}
 \begin{tabular}{lrrr}
  \hline \hline
 Isotopologue & $g_{\rm ns}$ & $N_{\rm states}$ & $N_{\rm trans}$\\
 \hline
   \pn & 6 & 30327 & 1333445 \\
  ${}^{31}$P${}^{15}$N & 4 & 31563 & 1438181 \\
 \hline \hline
\end{tabular}
\end{table}


\subsection{Energy levels}
The \Duo\ calculated energies were replaced with the MARVEL values (MARVELised), where available. We also used \textsc{PGOPHER} \citep{PGOPHER} to obtain Effective Hamiltonian (EH) energies to cover gaps in the MARVEL data of the \A\ state, using the constants provided by \citet{95AhHaxx.PN} and \citet{81GhVeVa.PN}. The \textsc{PGOPHER} file used is available as part of the supplementary material. In places where there were gaps in MARVEL energy levels in the range of $J=0\ldots30$ for \A\ state, we ran an analysis to determine whether \Duo\ or EH energies give better agreement with MARVEL within that range. Where EH energies were in better agreement with MARVEL energies, \Duo\ energies were substituted with EH energies. The summary of ranges of \Duo\ energies substituted by EH, where MARVEL energies were not available, is given in Table \ref{tab:pgopher_sub}, with only 26 \A\ rovibronic energy levels substituted overall.

\begin{figure}
    \centering
    \includegraphics[width=0.7\textwidth]{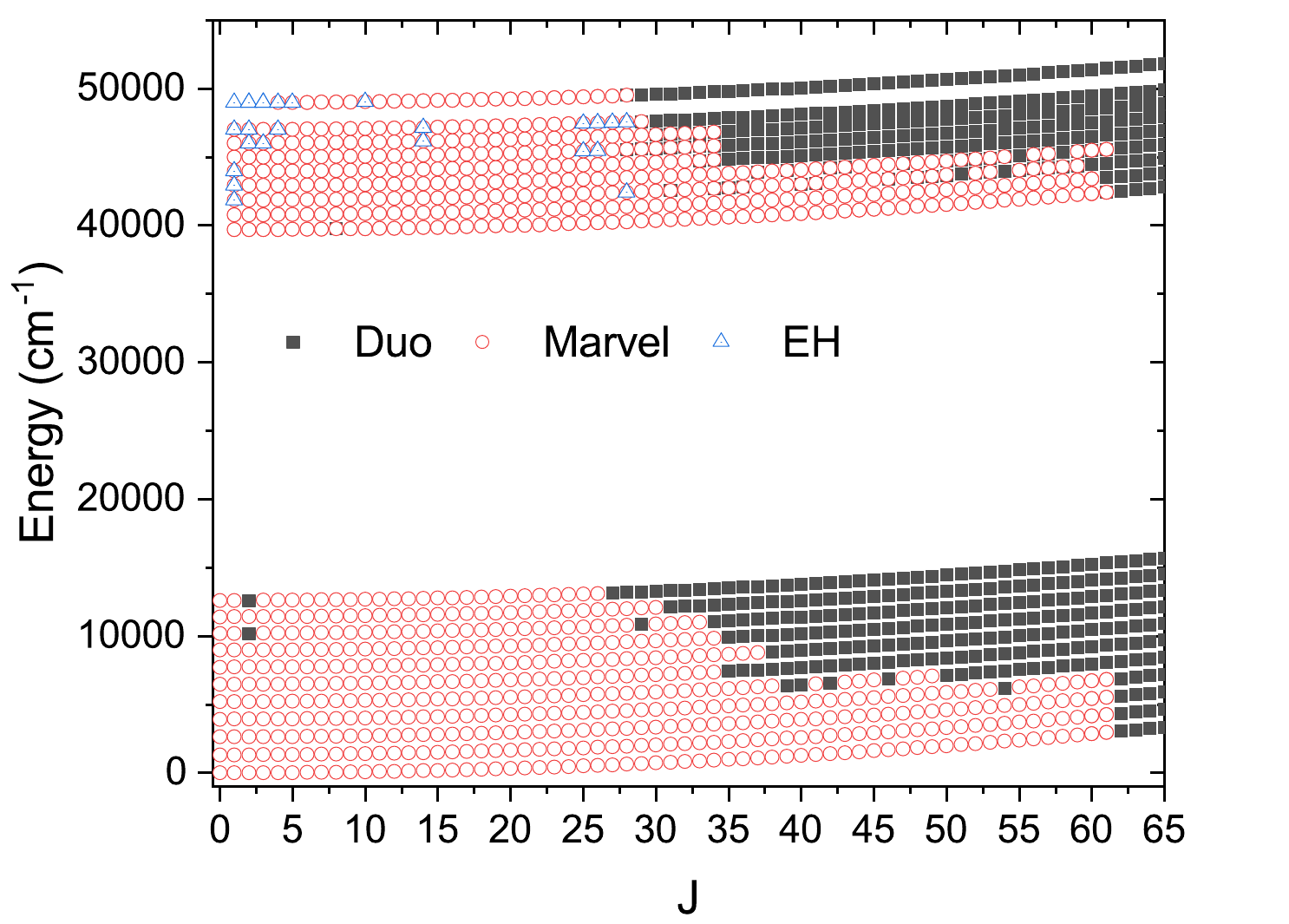}
    \caption{Visual representation of the sources of energy levels in the \name\ line list: \Duo\ values, MARVEL values and Effective Hamiltonian (EH) values for the vibrational states used in the MARVELisation procedure. The states shown in the graph represent only those with experimental data available, namely $v=0\dots11$ for \X\ and $v=0\dots7,9$ for \A.}
    \label{fig:marvelised_states}
\end{figure}

 We have used the labels `Ca', `EH' and `Ma' in the penultimate column of the States file to indicate if the energy value is calculated using \Duo, derived using Effective Hamiltonian (PGOPHER) or using MARVEL, respectively \citep{jt948}. It is worth noting that due to the perturbation of the \A\ rovibronic levels through near-lying dark states, we decided not to extrapolate MARVEL energy levels to higher values of $J$ using the predicted shift methodology as suggested by \citet{jt948} (as part of the QuadHybrid approach) as we could not guarantee the accuracy of extrapolated energy levels.  A visual representation of the sources of energy levels is given in Fig.~\ref{fig:marvelised_states} for the vibronic states affected by the MARVELisation.

\begin{table}
\caption{Description of the \A\ states, where the \Duo\ energies have been selected to be replaced either with  MARVEL or Effective Hamiltonian values in the MARVELisation procedure.}
\label{tab:pgopher_sub}
\centering
\begin{tabular}{lll}
\hline
 $v$ & Range of $J$'s substituted \\
\hline
 2 &  $J=1\cdots29$ \\
 3 &  $J=1\cdots29$ \\
 4 &  $J=1\cdots28$ \\
 5 &  $J=1\cdots26$ \\
 6 &  $J=2\cdots20$ \\
 7 &  $J=2\cdots28$ \\
 9 &  $J=1\cdots24$ \\
\hline
\end{tabular}

\label{table:decision_table}
\end{table}

The uncertainty values in the States file correspond to two cases: the MARVEL uncertainties are used for MARVELised energies, while for the \Duo\ and EH calculated values, the following approximate expression is used:
\begin{equation}
\label{e:unc}
\sigma(\text { state }, J, v)=\Delta T+\Delta \omega v+\Delta B J(J+1) \text {, }
\end{equation}
where $\Delta T$,  $\Delta \omega$ and $\Delta B$ are electronic state dependent constant, given in Table~\ref{t:unc:a:b}. For the \X\ and \A\ states, uncertainties were estimated based on the progression of residuals from Fig.~\ref{fig:marvelresiduals} as average increases of obs.-calc. in $v$ and  $J$ for each state shown.  The uncertainties for the levels affected by the perturbations were increased according to the corresponding MRSEs.

\begin{table}
\centering
\caption{Constants $\Delta T$, $\Delta \omega$ and $\Delta B$ (\cm) defining state dependent uncertainties via Eq. \eqref{e:unc} for `Ca' and `EH' energy levels.}
\label{t:unc:a:b}
\begin{tabular}{llll}
\hline State & $\Delta T$ & $\Delta \omega$ & $\Delta B$ \\
\hline \X\  & 0.057 & 0.05 & 0.0001 \\
       \A\  & 0.237 &  0.05 & 0.0001 \\
\hline
\end{tabular}
\end{table}

\subsection{Partition Function}

The partition function (PF) is an important tool when linking the energy levers of a molecule to its macroscopic properties, enabling predictions in astrophysical and atmospheric contexts. As such, as part of this work, we have computed PF for \pn\ and $^{31}$P$^{15}$N as part of our semi-empirical line list. The function was computed using:
\begin{equation}
\label{e:PF}
Q(T)=\sum_i g_i^{\text {tot }} \mathrm{e}^{-\frac{c_2 \Tilde{E}_i}{T}},
\end{equation}
where $c_2$ is the second radiation constant, $T$ is temperature in Kelvin, $\Tilde{E}_i$ is the total rovibronic energy in \cm\ of a state $i$ and $g_i^{\text {tot }}$ is the total state degeneracy given by
\begin{equation}
    \label{e:gntot}
    g_i^{\text {tot }}=g_{\text {ns }}(2J_i+1),
\end{equation}
where $g_{\text {ns}}$ is the nuclear weight spin-statistic. ExoMol uses the HITRAN convention, represented by the
full nuclear statistical weight, meaning that for \pn\ $g_{\text {ns}}$ = 6 and for ${}^{31}$P${}^{15}$N $g_{\text {ns}}$ = 4, unlike the standard astrophysics convention, where $g_{\text {ns}}$ is not included in PF.

\begin{figure}
    \centering
    \includegraphics[width=0.7\textwidth]{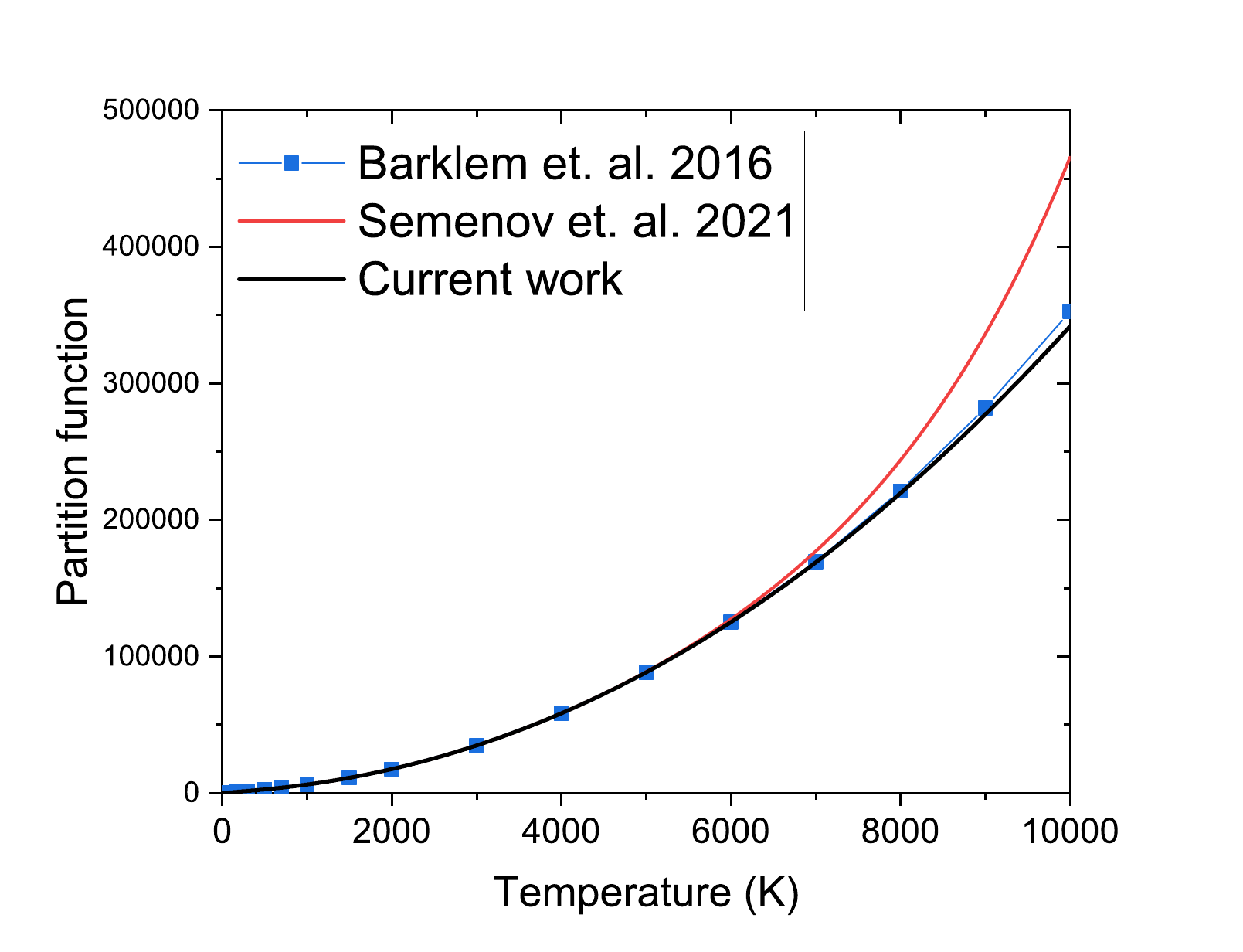}
    \caption{Comparison of partition functions between current work, previous work \citet{jt842} and that of \citet{16BaCoxx.partfunc}.}
    \label{fig:PF}
\end{figure}

Figure \ref{fig:PF} compares the PF between our current work, our previous \ai\ work \citep{jt842} and that of \citet{16BaCoxx.partfunc}. We calculate our PF in a similar procedure used previously in \citet{jt842} using a  1 K temperature step. For comparison of the \pn\ PF with \citet{16BaCoxx.partfunc}, we multiply their values by a factor of six to account for the lack of nuclear weight spin-statistic in their calculations, which we include. Our PF is in good agreement with previous results up to 5000~K but is then lower than the results of \citet{jt842}. The differences at higher temperatures can be explained by the additional electronic states included in the \ai\ study \citep{jt842}. We fix the maximum temperature of
PaiN to be 5~000~K.

\subsection{Lifetimes}
\label{s:lifetimes}

As part of the \name\ line list, we provide lifetimes for the \X\ and \A\ states in our model. The lifetimes are calculated using ExoCross for each unique rovibronic energy level using the .states and .trans files. ExoCross uses the following equation to calculate lifetimes:
\begin{equation}
\label{e:lifetimes}
      \tau_{i}=\frac{1}{\sum_{j<i}{A_{ij}}}
\end{equation}
where $\tau_i$ is radiative lifetime, $A_{ij}$ is Einsteins A coefficients, and $i$ and $j$ stand for upper and lower states, respectively.

\begin{figure}
    \centering
    \includegraphics[width=0.7\textwidth]{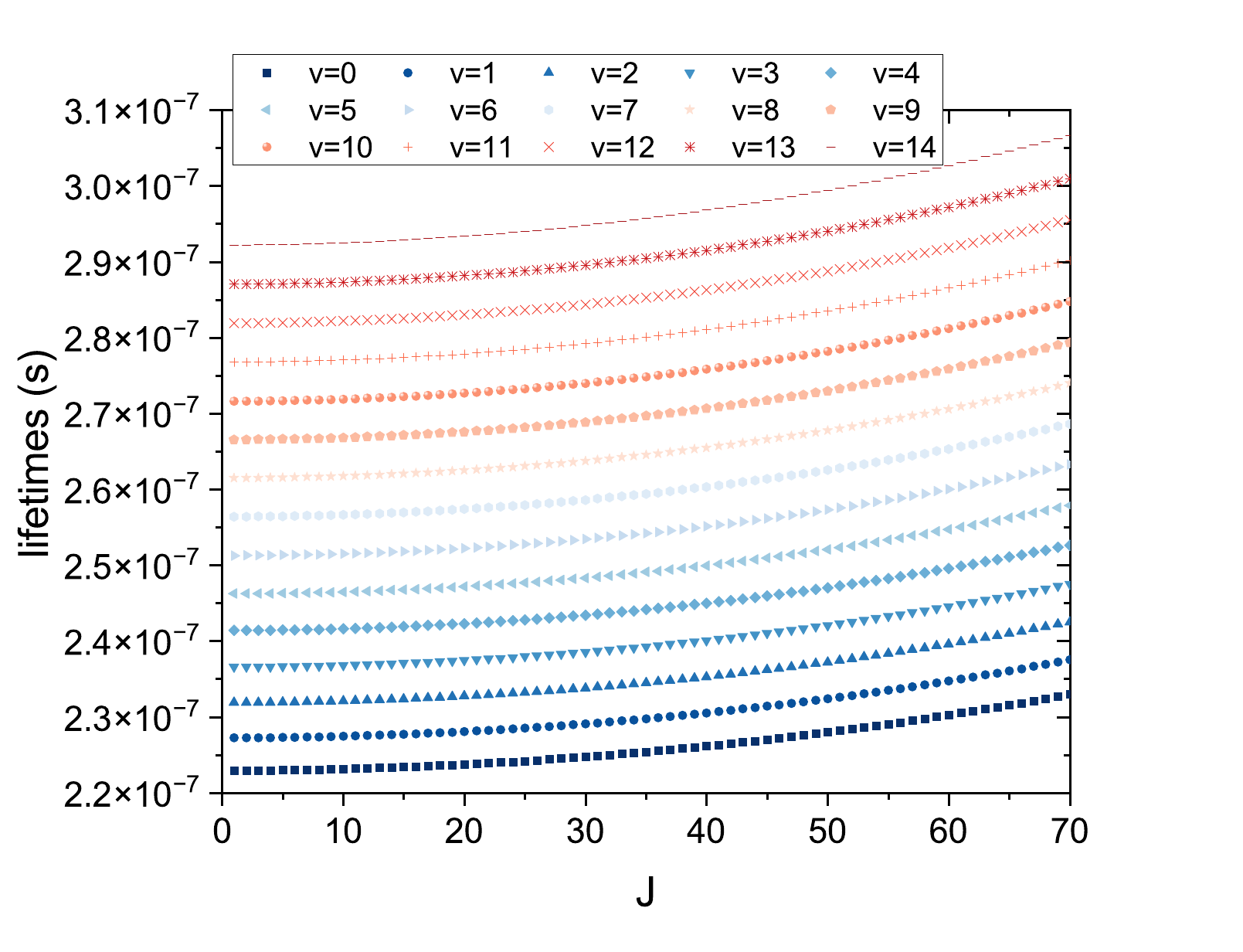}
    \caption{Lifetimes of the \A\ state. }
    \label{fig:lifetimes}
\end{figure}

There is only one experimental measurement of the lifetime of \pn\ for the \A\ $v=0$ state, and it is equal to 227 $\pm$ 70 ns. We used this lifetime to scale our \XA\ transition dipole by a factor of 1.28 to get within reasonable accuracy. Table \ref{tab:lifetimes} provides a comparison of the lifetimes produced by our scaled dipole, and previous works (\citet{jt842}, \citet{19QiZhLi.PN}). Figure \ref{fig:lifetimes} shows the lifetimes for $J=0 \dots 70$ rotational states of the first 14 vibrational levels of \A. From the figure, we can see that there is little to no $J$ dependence in the \A\ lifetimes. It is also worth noting that because we only included two electronic states in our model, the lifetimes appear unperturbed which would not be the case should we include adjacent electronic states for \A.

\begin{table}
\caption{Comparison of lifetimes from (A) our current work, (B) \citet{jt842} and  (C) \citet{19QiZhLi.PN}. An experimental lifetime 227 $\pm$ 70 ns was reported by  \citet{75MoMcSi.PN} for $v'=0$ of \A, which is in good agreement with our results.}
\label{tab:lifetimes}
\begin{tabular}{rlll}
    \hline
    \hline
   & \multicolumn{3}{c}{\A\ /ns} \\
    \hline
$v'$ & A & B & C \\
    \hline
0   & 222.9 & 341.3 & 659.4 \\
1   & 227.2 & 334.6 & 674.3 \\
2   & 231.9 & 338.9 & 660.8 \\
3   & 236.6 & 344.3 & 646.5 \\
4   & 241.4 & 349.8 & 642.2 \\
5   & 246.3 & 356.2 & 643.0 \\
6   & 251.3 & 361.6 & 632.1 \\
7   & 256.4 & 368.1 & 610.6 \\
8   & 261.5 & 376.3 & 607.4 \\
9   & 266.6 & 392.4 & 598.6 \\
10  & 271.6 & 399.3 & 583.8 \\
11  & 276.8 & 407.3 & 581.8 \\
12  & 281.9 & 413.0 & 566.8 \\
13  & 287.1 & 420.7 & 565.3 \\
14  & 292.2 & 431.3 & 553.1 \\
    \hline
    \hline
\end{tabular}
\end{table}

\subsection{Line lists and opacities}
The line list is provided in State and Transition files, as is customary for the ExoMol format \citep{jt548,jt939}.  Extracts from the States and Trans files are shown in Tables \ref{t:states} and \ref{t:transfile}, respectively; the full files are available from \url{www.exomol.com}. The States file contains state IDs, energy term values (\cm), state uncertainties (\cm), Land\'{e}-$g$ factors \citep{jt656}, lifetimes (s) \citep{jt624} and quantum numbers. The Transition file contains the state IDs and Einstein A coefficients (s$^{-1}$). The partition functions are also included in the standard line list compilation on a grid of 1~K between 0 and 10~000~K.

\begin{table}
\centering
\caption{ Extract from the \name\ .state file }
\label{t:states}
{\tt  \begin{tabular}{rrrrrlrlllrrrrcr} \hline \hline
$i$ & \multicolumn{1}{c}{$\tilde{E}$ (\cm)} & \multicolumn{1}{c}{$g_i$} & \multicolumn{1}{c}{$J$} & \multicolumn{1}{c}{unc. (\cm)} & \multicolumn{1}{c}{$\tau$ (s$^{-1}$)} &  \multicolumn{1}{c}{g} & \multicolumn{2}{c}{Parity} 	& \multicolumn{1}{c}{State}	& $v$	&${\Lambda}$ &	${\Sigma}$ & $\Omega$ & \multicolumn{1}{c}{Label} & \multicolumn{1}{c}{$\tilde{E}$ (\cm)} \\
\hline
31 & 39690.318649 & 18 & 1 & 0.198704 & 2.2294E-07 & 0.500000 & + & f & A1Pi & 0 & 1 & 0 & 1 & Ma & 39690.342964 \\
32 & 40778.595433 & 18 & 1 & 0.100000 & 2.2726E-07 & 0.500000 & + & f & A1Pi & 1 & 1 & 0 & 1 & Ma & 40778.599149 \\
33 & 41852.997294 & 18 & 1 & 0.759055 & 2.3190E-07 & 0.500000 & + & f & A1Pi & 2 & 1 & 0 & 1 & Ca & 41852.997294 \\
34 & 42912.839136 & 18 & 1 & 0.809055 & 2.3664E-07 & 0.500000 & + & f & A1Pi & 3 & 1 & 0 & 1 & Ca & 42912.839136 \\
35 & 43957.866467 & 18 & 1 & 0.859055 & 2.4142E-07 & 0.500000 & + & f & A1Pi & 4 & 1 & 0 & 1 & Ca & 43957.866467 \\
36 & 44989.841525 & 18 & 1 & 0.700000 & 2.4626E-07 & 0.500000 & + & f & A1Pi & 5 & 1 & 0 & 1 & Ma & 44988.062552 \\
37 & 46002.911525 & 18 & 1 & 0.500000 & 2.5128E-07 & 0.500000 & + & f & A1Pi & 6 & 1 & 0 & 1 & Ma & 46003.572725 \\
38 & 47004.651492 & 18 & 1 & 1.009055 & 2.5642E-07 & 0.500000 & + & f & A1Pi & 7 & 1 & 0 & 1 & Ca & 47004.651492 \\
39 & 47991.618462 & 18 & 1 & 1.059055 & 2.6151E-07 & 0.500000 & + & f & A1Pi & 8 & 1 & 0 & 1 & Ca & 47991.618462 \\
\hline
\hline
\end{tabular}}
\mbox{}\\
{\flushleft
$i$:   State counting number.     \\
$\tilde{E}$: State energy term values in \cm, MARVEL, EH  or Calculated (\textsc{Duo}). \\
$g_i$:  Total statistical weight, equal to ${g_{\rm ns}(2J + 1)}$.     \\
$J$: Total angular momentum.\\
unc: Uncertainty, \cm.\\
$\tau$: Lifetime (s$^{-1}$).\\
$g$: Land\'{e} $g$-factors. \\
$+/-$:   Total parity. \\
State: Electronic state.\\
$v$:   State vibrational quantum number. \\
$\Lambda$:  Projection of the electronic angular momentum. \\
$\Sigma$:   Projection of the electronic spin. \\
$\Omega$:   Projection of the total angular momentum, $\Omega=\Lambda+\Sigma$. \\
Label: `Ma' is for MARVEL, `EH' is for Effective Hamiltonian and `Ca' is for Calculated. \\
$\tilde{E}$: State energy term values in \cm, Calculated (\textsc{Duo}). \\
}
\end{table}

\begin{table}
\centering
\caption{Extract from the .trans file of the \name\ line list for  \pn.}
\tt
\label{t:transfile}
\centering
\begin{tabular}{rrrr} \hline\hline
\multicolumn{1}{c}{$f$}	&	\multicolumn{1}{c}{$i$}	& \multicolumn{1}{c}{$A_{fi}$ (s$^{-1}$)}	&\multicolumn{1}{c}{$\tilde{\nu}_{fi}$} \\ \hline
    77 & 6 & 1.0583E-10 &    13009.188326 \\
    78 & 6 & 1.2843E-11 &    14104.631827 \\
    79 & 6 & 3.0917E-12 &    15185.086622 \\
    80 & 6 & 1.4723E-12 &    16250.408774 \\
    81 & 6 & 3.0994E-13 &    17300.443346 \\
    82 & 6 & 1.2750E-14 &    18335.023853 \\
    83 & 6 & 1.7852E-14 &    19353.971649 \\
    84 & 6 & 2.2823E-14 &    20357.095245 \\
    85 & 6 & 1.3905E-14 &    21344.189551 \\
    \hline\hline
\end{tabular} \\ \vspace{2mm}
\rm
\noindent
$f$: Upper  state counting number;\\
$i$:  Lower  state counting number; \\
$A_{fi}$:  Einstein-$A$ coefficient in s$^{-1}$; \\
$\tilde{\nu}_{fi}$: transition wavenumber in \cm.\\
\end{table}

Temperature- and pressure-dependent molecular opacities of \pn\ based on the \name\ line list have been generated using the ExoMolOP procedure~\citep{jt801} for four exoplanet atmospheric retrieval codes: ARCiS~\citep{ARCiS}, TauREx~\citep{TauRexIII}, NEMESIS~\citep{NEMESIS} and petitRADTRANS~\citep{19MoWaBo.petitRADTRANS}. For the line broadening, we assumed an 86\% H$_2$ and 14\% He atmosphere and Voigt line profile.

\begin{figure}
    \centering
    \includegraphics[width=0.45\textwidth]{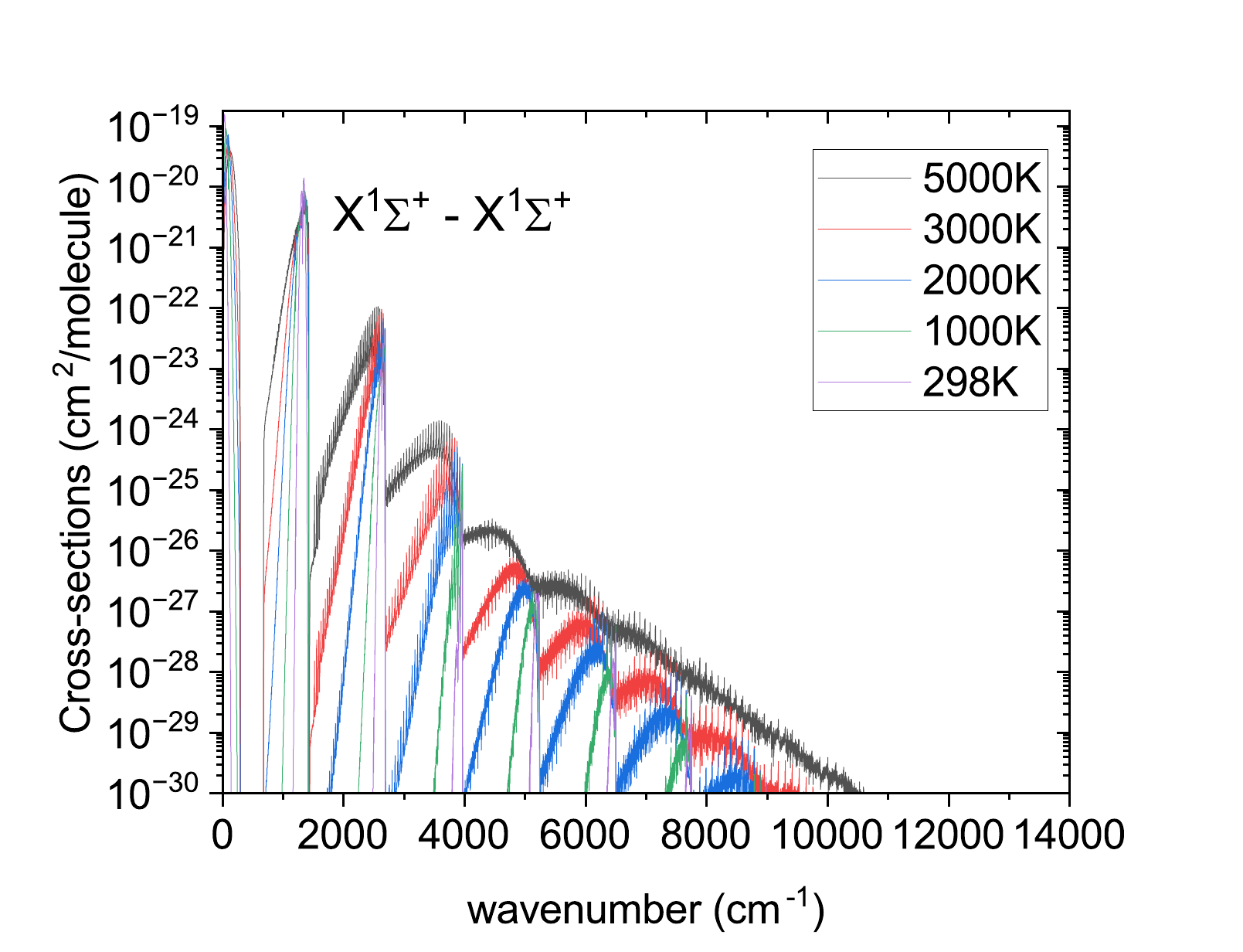}
    \includegraphics[width=0.45\textwidth]{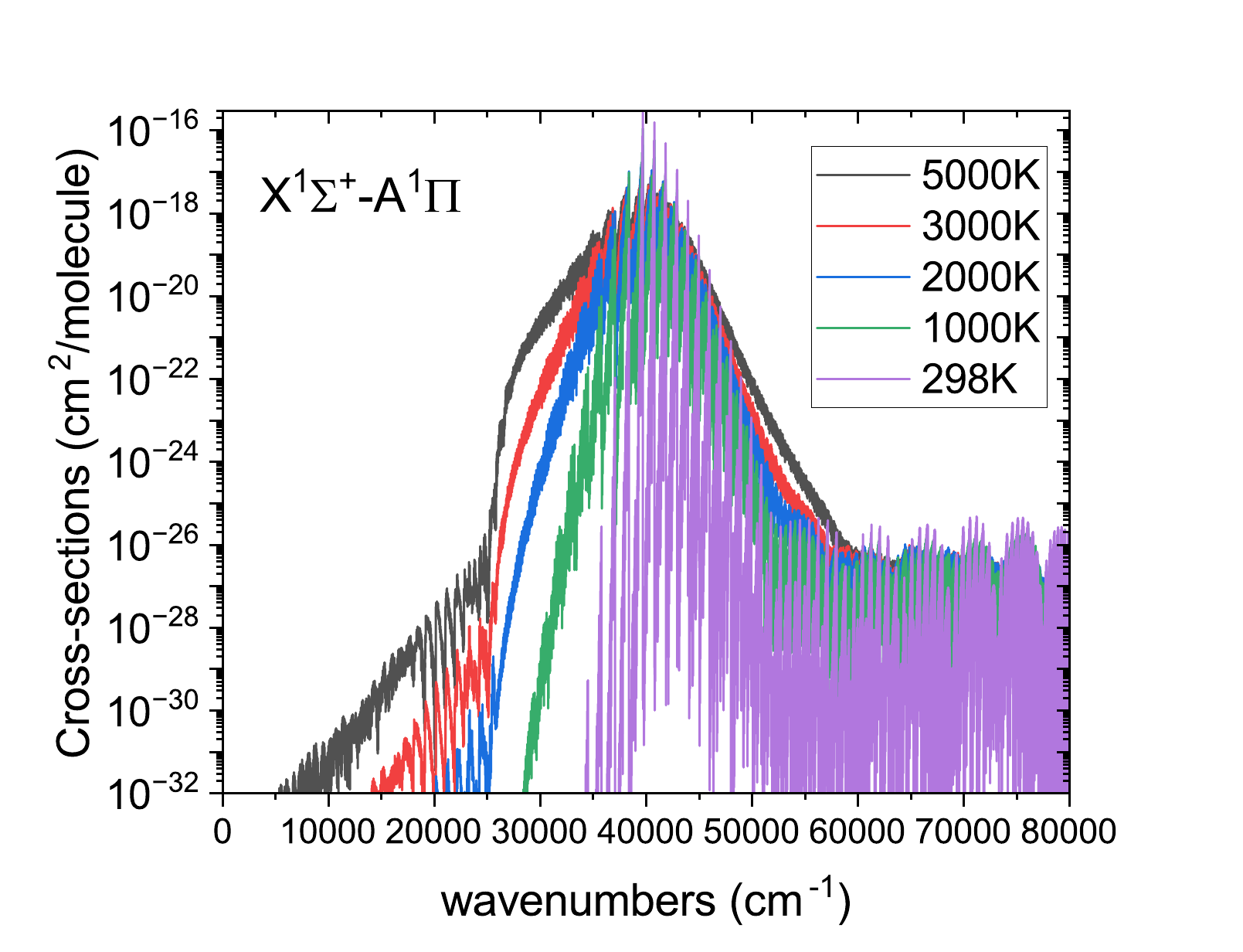}
    \caption{Overview of simulated \XA\ and \XX\ absorption spectra, both calculated at different temperatures with a Gaussian profile of HWHM = 1 \cm.}
    \label{fig:XAoverall}
\end{figure}

An overview of the simulated spectra of the \XA\ and \XX\ bands at different temperatures is given in Fig.~\ref{fig:XAoverall}. Owing to the specially constructed permanent \X\ DMC \citep{23UsSeYu.PN} used, the intensities of the \X\ overtone bands drop exponentially with no nonphysical plateau-lie features.

\subsection{The \XA\ band system}

\begin{figure}
    \centering
    \includegraphics[width=0.7\textwidth]{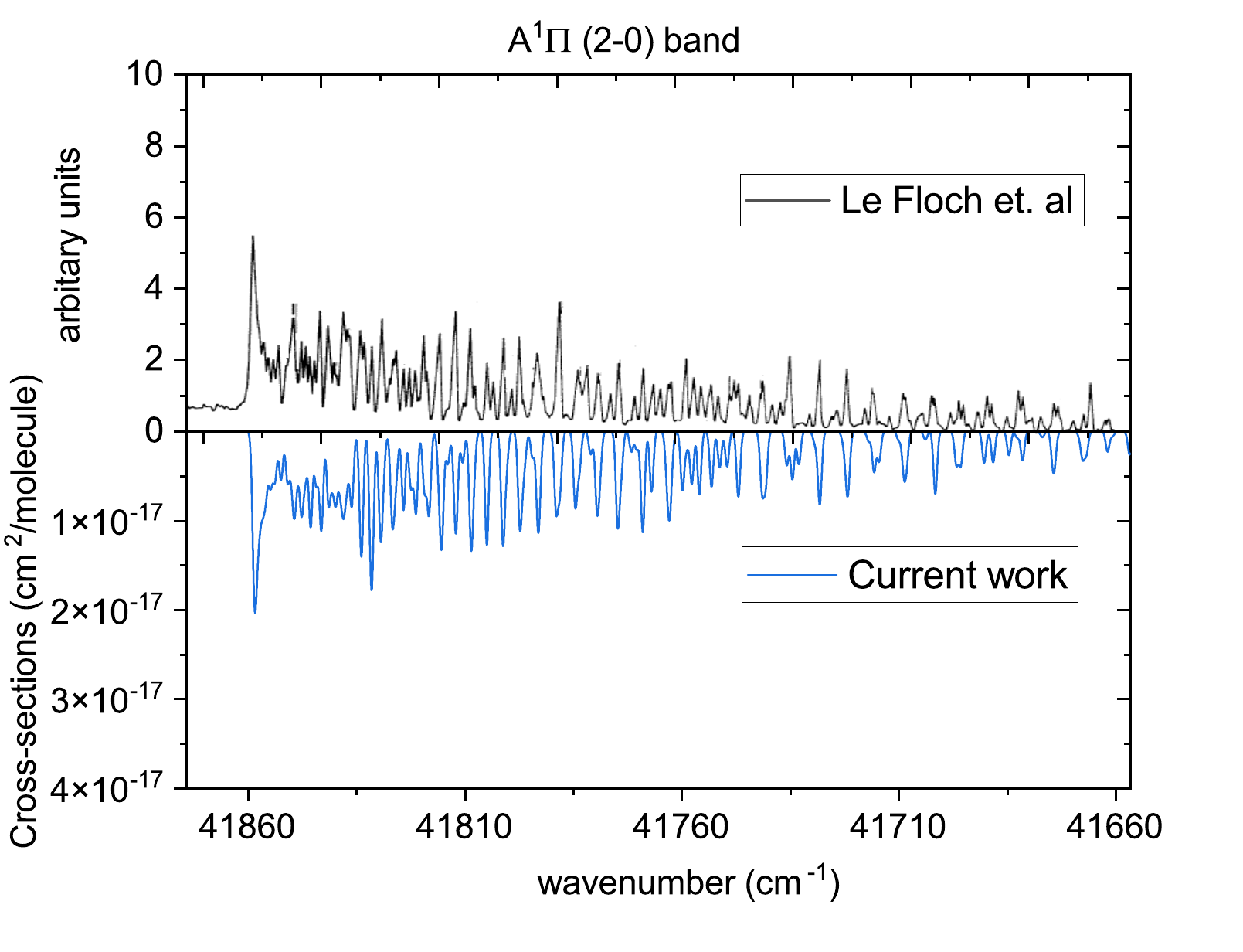}
    \caption{Comparison of the \XA\ (2,0) simulated absorption band (lower panel) with that recorded by \citet{96LeMeDu.PN} (upper panel). The spectrum was simulated at $T$ =1173.15 K with a Gaussian profile of HWHM = 0.5 \cm.}
    \label{fig:96LeMeDu}
\end{figure}

Originally discovered by \citet{33CuHeHe.PN},  the \XA\ band system has proved to be difficult to study due to the many perturbations from the near-lying electronic states as discussed in section \ref{s:perturbations} and shown previously experimentally by \citet{96LeMeDu.PN} and \citet{81GhVeVa.PN}. 
Below, we present simulated spectra of this band and compare them with observed results. Figure \ref{fig:96LeMeDu} provides a comparison of the rovibronic absorption spectrum of the \XA\ (2,0) band between the simulated spectrum and observations reported by \citet{96LeMeDu.PN}. As can be seen, the simulated spectrum accurately reproduces the experimental observations even for the perturbed $v=2$ state in \A. There is a notable improvement of 2217 \cm\ shift in comparison to our previous work \citep{jt842}, which is achieved by fitting the PEC to experimental data and using MARVELsation. The absorption spectrum was simulated with a Gaussian profile of HWHM = 0.5 \cm\ at a temperature of 1173.15~K which was extracted from the experimental work.

\begin{figure}
    \centering
    \includegraphics[width=0.5\textwidth]{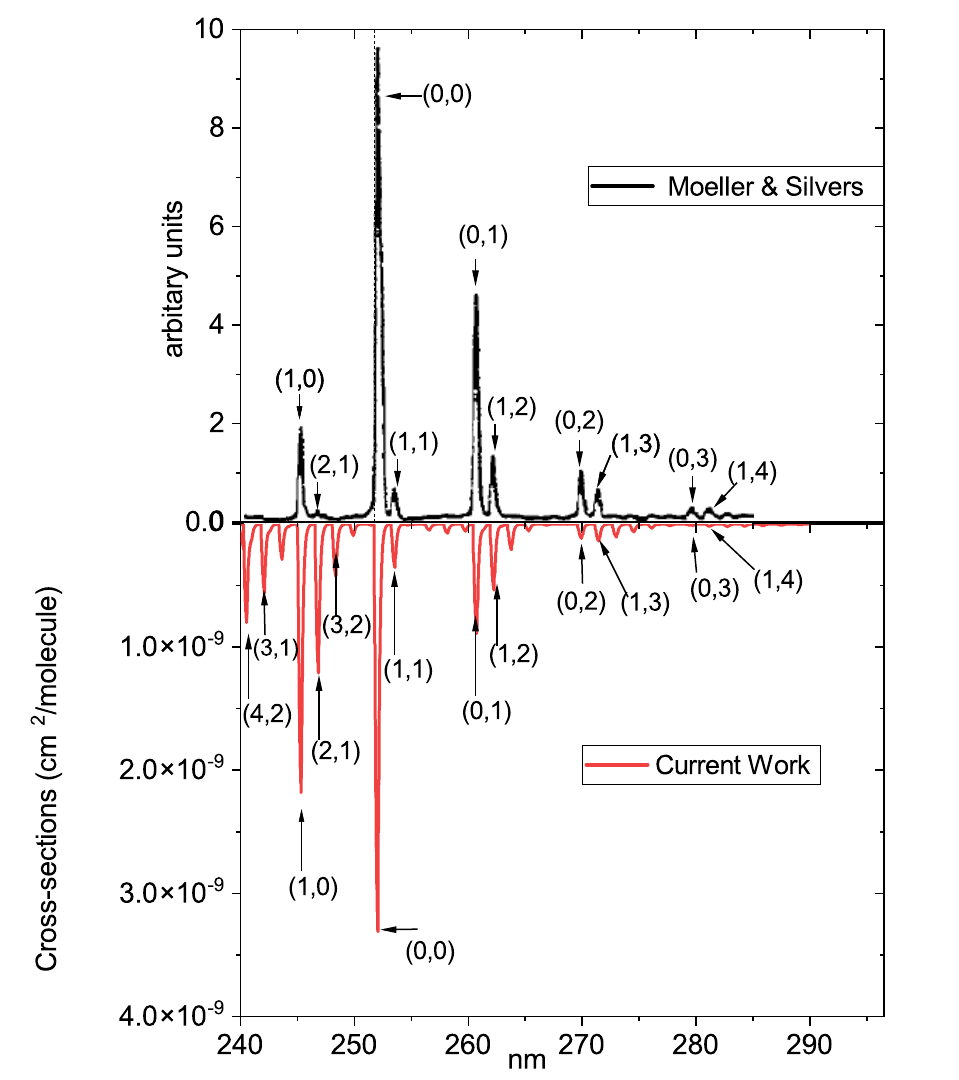}
    \caption{Comparison of an \XA\ simulated emission spectrum (lower panel) with that recorded by \citet{73MoSixx.PN} (upper panel). The numbers in brackets represent the vibrational bands. The spectrum was simulated at $T_{\rm rot}=$ 300~K (rotational) and $T_{\rm vib}=$ 3500~K (vibrational) temperatures with a Gaussian profile of HWHM = 100 \cm\ . }
    \label{fig:73MoSixx}
\end{figure}

Figure \ref{fig:73MoSixx} offers an illustration of the  \XA\  emission spectrum calculated using \name\ and compared to a non-LTE (non Local Thermal Equilibrium) emission spectrum reported by \citet{73MoSixx.PN}. As an attempt to describe the complex non-LTE environment of the experiment, a simple, two-temperature non-LTE model \citep{19PaLaxx} was used as implemented in ExoCross \citep{ExoCross}, with $T_{\rm rot}=$ 300~K (rotational) and $T_{\rm vib}=$ 3500~K (vibrational) temperatures.  The figure also highlights vibrational bands both in the experimental and simulated spectra. Some bands in the simulated spectra are still too weak to be seen, namely (0,2), (1,3), (0,3), (3,1), and would require a more sophisticated non-LTE model for better agreement.

\section{Conclusions}

New line lists for \pn\ and $^{31}$P$^{15}$N are presented, covering wavenumbers in the range from 0 to 82~500 \cm. The \name\ line lists supersede the original PN ExoMol line lists YYLT \citep{jt590} by extending it to the \XA\ band.  \name\ is available from \url{www.exomol.com}.

As part of the line list update, a MARVEL analysis for \pn\ was performed using 6005 measured rovibronic transitions. All experimental line positions from the literature (to the best of our knowledge) covering the IR and UV regions of \XX\ and \XA\ systems were collected and processed to generate a comprehensive, self-contained set of empirical energies of \pn. An accurate spectroscopic model for PN in \X\ and \A\
was built using previously calculated \ai\ TDMCs and empirically refined PECs, EAMC and a $\Lambda$-doubling curve.

The line list for \pn\ was MARVELised, where the \Duo\ calculated energies were replaced with the MARVEL or EH values (where available and applicable). The \name\ line list provides uncertainties of the rovibronic states in order to help in high-resolution applications.  Comparisons of simulated spectra show close agreement with the experiment. By scaling the \AX\ TDMC  an agreement between computed reported lifetimes has been achieved.

The new line lists  should help detect  PN in the UV region and be useful in atmospheric chemistry applications.  Further improvements to the line lists can include adding the dark states to the model to accurately model the \A\ perturbations and the expansion of the line list to include the \E\ state. Both of these improvements will require further \ai\ calculations coupled with additional experimental data not currently available.

\section*{Acknowledgements}

This work was supported by the European Research Council (ERC) under the European Union's Horizon 2020 research and innovation programme through Advance Grant number 883830, and  STFC Project No. ST/Y001508/1.
N.E-K's work is supported by ASPIRE AARE grant number AARE20-000329-00001  and Khalifa University of Science and
Technology grant 8474000336-KU-SPSC.

\section{Data Availability} 
The data underlying this article are available in the article and in its online supplementary material. The line list and associated data (partition functions, opacities, temperature-dependent absorption cross sections) for PN are available from \href{www.exomol.com}{www.exomol.com}. The codes used in this work, namely \textsc{Duo} and \textsc{ExoCross}, are freely available via \href{https://github.com/exomol}{https://github.com/exomol}.

\section*{Supporting Information}
The following is provided as supporting information (1) the \textsc{Duo} input file which contains all the PN potential energy, dipole moment and coupling curves used in this work and (2) the MARVEL input (transitions and segment) file and
output (energy) file, (3) PGOPHER file with the effective Hamiltonian constants used in this work and (4) documented transitions by \citet{87VeGhIq.PN}.







\bsp	
\label{lastpage}
\end{document}